\documentclass[fleqn,usenatbib]{mnras}

\usepackage{newtxtext,newtxmath}

\usepackage[T1]{fontenc}

\DeclareRobustCommand{\VAN}[3]{#2}
\let\VANthebibliography\thebibliography
\def\thebibliography{\DeclareRobustCommand{\VAN}[3]{##3}\VANthebibliography}

\usepackage{graphicx}	
\usepackage{amsmath}	
\usepackage[figuresright]{rotating}
\usepackage{gensymb}
\usepackage{natbib}
\usepackage{multicol}

\title[Unravelling the Period Gap]{Unravelling the Period Gap using LAMOST Chromospheric Activity Indices}

\author[D. Chahal et al.]{
Deepak Chahal$^{1,2}$\thanks{E-mail: deepakchahal294@gmail.com},
Devika Kamath$^{1,2,3}$,
Richard de Grijs$^{1,2}$,
Paolo Ventura$^{3,4}$
and Xiaodian Chen$^{5}$
\\
$^{1}$School of Mathematical and Physical Sciences, Macquarie University, Balaclava Road, Sydney, NSW 2109, Australia\\
$^{2}$Astrophysics and Space Technologies Research Centre, Macquarie University, Balaclava Road, Sydney, NSW 2109, Australia\\
$^{3}$INAF, Observatory of Rome, Via Frascati 33, 00077 Monte Porzio Catone, Italy \\
$^{4}$Istituto Nazionale di Fisica Nucleare, section of Perugia, Via A. Pascoli snc, 06123 Perugia, Italy \\
$^{5}$CAS Key Laboratory of Optical Astronomy, National Astronomical Observatories, Chinese Academy of Sciences, Beijing 100101, China
}

\date{Accepted 2023 August 15. Received 2023 July 28; in original form 2023 April 19}

\pubyear{2023}

\begin{document}
\label{firstpage}
\pagerange{\pageref{firstpage}--\pageref{lastpage}}
\maketitle

\begin{abstract} 
In our recent catalogue of BY Draconis (BY Dra) variables based on Zwicky Transient Facility data, we found traces of a period gap in the period--colour diagram. We combined our BY Dra database with catalogues from the {\sl Kepler} and K2 surveys, revealing a prominent period gap. Here, we use this combined ZTF--{\sl Kepler}--K2 data set to investigate the origin of the period gap observed for BY Dra stars using chromospheric activity indices. We use low- and medium-resolution spectra from the LAMOST Data Release 7 to derive magnetic activity indices for the Ca {\sc ii} H and K and H$\alpha$ emission lines. We find a strong dependence of chromospheric activity on both stellar mass and rotation period. For partially convective K--M-type stars, the activity decreases steeply up to an age of $\sim$700--1000 Myr, subsequently evolving to the type of low-level saturation associated with spin-down stallation. In  contrast, F--G-type stars with thinner convective envelopes exhibit constant activity with increasing age. We suspect that the observed steep decrease for partially convective stars is driven by core--envelope coupling. This mechanism reduces differential rotation at the core--envelope transition, hence leading to decreased magnetic activity. Moreover, we derive activity indices for previously known star clusters and find similar trends as regards their activity levels as a function of age. In particular, very low-level activity is observed around the location of the period gap. Therefore, we conclude that the period gap, defined by the non-detection of variable sources, is driven by a minimum in chromospheric activity. 
\end{abstract}





\section{Introduction} \label{sec:section1}

Magnetic-field regeneration occurs in solar-type stars through the dynamo process and leads to phenomena such as large and rapidly evolving starspots (cooler regions where convection is suppressed by magnetic-field lines) in the photosphere, strong Ca {\sc ii} and Balmer emission lines from the chromosphere and prominent flares or superflares (sudden releases of magnetic energy) from the corona, among others \citep[][and references therein]{berdyugina2005starspots,brun2015solar,deGrijs2021}. Dark starspots move across a stellar disc through rotation, giving rise to the gradual modulation of a star’s luminosity \citep{berdyugina2005starspots,balona2019rotational}. Such stars are generally referred to as rotational modulators or rotational variables. Rotational variables are also characterised by non-thermal emission in the cores of certain chromospheric lines such as Ca {\sc ii} H and K \citep[Mount Wilson Program;][]{Wilson1968,duncan1991ii}, Mg {\sc ii} \citep{Hall2008} and H$\alpha$ \citep{fang2016stellar,fang2018stellar,zhang2019stellar}, which are widely used as indicators of surface magnetic activity in cool stars.

Magnetic-field regeneration, especially in Sun-like stars, occurs at the tachocline, the interface between a star's internal radiative and convective zones \citep[][and references therein]{charbonneau1999helioseismic,brun2015solar,deGrijs2021}. Stellar winds co-rotate with the stellar surface out to the Alfv\'en radius, where they decouple and angular momentum is lost. Therefore, the strength of the magnetic field and the angular momentum lost depend on the stellar rotation period. For that reason, it has been thought that although stars are born with a range of rotation periods, rapidly rotating stars lose angular momentum quickly and converge onto a well-defined, slowly rotating sequence \citep{spada2020competing}. Several studies have found a dearth of sources at certain rotation periods, depending on stellar temperature, in statistically complete data sets \citep{Mcquillan2014,Reinhold2020,Gordon2021}. The resulting low density of data points in the colour--period diagram has become known as the `period gap'. It is associated with the slowly rotating sequence (stars with $P_\textrm{rot} >$10--15 days) in the diagnostic period--colour diagram. The gap starts at a period of $\sim$20 days for G dwarfs and extends up to $\sim$30 days for early-M dwarfs \citep{Gordon2021,lu2022bridging}. The reality of such a gap is not obvious for stars more massive than K dwarfs.

Several explanations have been put forward to understand the gap's origin. The most frequently discussed hypothesis is a temporary epoch of spin-down stallation, followed by a rapid spin-down \citep{Gordon2021}. FGKM-type main-sequence stars lose mass through magnetised stellar winds, which leads to a decrease in spin so as to conserve angular momentum. Recent measurements of rotation periods in benchmark open clusters such as Praesepe (670 Myr), the Hyades (730 Myr), NGC 6811 (1 Gyr), and NGC 752 (1.4 Gyr) have shown that the rate of spin-down is mass-dependent, and it is also halted temporarily around the age of the Praesepe cluster \citep{van2016weakened,agueros2018new,curtis2019temporary,curtis2020stalled,david2022further}. It has been found that, after converging onto a slowly rotating sequence, stars temporarily stop spinning down. The subsequent rapid spin-down produces a physical gap in the period--colour diagram. The temporarily stalled spin-down is thought to be caused by the onset of angular momentum transport from the radiative core to the convective envelope \citep{spada2020competing}, which is referred to as core--envelope coupling. Recently, \citet[][using data from the Zwicky Transient Facility; ZTF]{lu2022bridging} showed that the period gap closes at the fully convective limit. This supports the core--envelope coupling hypothesis.

To complete the picture of stellar evolution, stellar rotational evolution must be constrained. In this paper, we derive chromospheric activity indices from Ca {\sc ii} H and K and H$\alpha$ line emission. We aim to study the origin of the period gap by exploring the behaviour of a number of activity indices along the slowly rotating sequence. The exact mechanism that drives the stellar dynamo process and, hence, leads to the regeneration of the magnetic field over different spectral types is not yet properly understood. Therefore, we also aim to investigate the behaviour of the rotation--age, activity--age and rotation--activity relationships for different stellar spectral types, given that none of them are as yet fully constrained \citep{barnes2007ages,mamajek2008improved}. A secondary objective is to understand how chromospheric activity evolves over time for different temperature ranges. 

In our previous study \citep{chahal2022statistics}, we found traces of a period gap at higher stellar masses for BY Draconis (BY Dra) variable stars in the ZTF catalogue. BY Dra stars are FGKM-type main-sequence stars that show starspot modulation in their luminosities. They represent a special class of rotational variables which show strong emission lines in their chromospheres. To undertake a complete statistical study of their activity, we collected complementary data from previously published databases which also reveal the presence of a period gap. In Section \ref{sec:section2} we present our target sample and provide a brief description of the spectra we have used. Section \ref{sec:section3} offers an overview of the techniques used to derive activity indices from our target objects' Ca {\sc ii} H and K and H$\alpha$ line emission. In Section \ref{sec:section4} we present an analysis of the behaviour of the chromospheric activity indices as a function of a number of different stellar parameters. We also analyse the evolution of active stars using the rotation--colour diagram. In Section \ref{sec:section5} we discuss evidence of spin-down stallation and its implications. We conclude the paper in Section \ref{sec:section6}.

\section{Data and Observations} \label{sec:section2}

This is a detailed follow-up study using our previously published catalogue of BY Dra variables \citep{chahal2022statistics}. In this section, we discuss our target sample, which was obtained by combining data from multiple catalogues, and their corresponding spectra.

\subsection{Target Sample}

We have enlarged our target sample by gathering data of BY Dra rotational variables from the {\sl Kepler} \citep{Mcquillan2014} and K2 surveys \citep{Reinhold2020}, along with our ZTF BY Dra catalogue of $\sim$78,000 stars \citep{chahal2022statistics}. These catalogues also exhibit traces of a period gap. Using the data thus collected, we found cross-matches of $\approx$15,000 spectra in the Large Sky Area Multi-Object Fibre Spectroscopic Telescope (LAMOST; the Guo Shoujing Telescope) low-resolution DR7 database; with the LAMOST medium-resolution data, the cross-match was limited to $\approx$4000 spectra. The combined catalogue contains a wide range of rotation periods (1--40 days) and effective temperatures ($T_\textrm{eff} = 3500$--6500 K).

The spectra, obtained with the Guo Shoujing Telescope \citep{cui2012large}, cover the relevant chromospheric indicators (Ca {\sc ii} H and K, H$\alpha$). Therefore, we used the LAMOST Data Release 7 (DR7) low- and medium-resolution spectra to estimate our sample stars' chromospheric activity. The Guo Shoujing Telescope is a reflecting Schmidt telescope equipped with 4000 fibres covering a field of view of 20 deg$^{2}$. LAMOST has collected millions of stellar spectra in service mode. The LAMOST low-resolution spectra (LAMOST LRS) cover the wavelength range of 3690--9100 {\AA} with a spectral resolution of $R = 1800$ at 5500 {\AA}. The facility's medium-resolution spectra (LAMOST MRS) have been taken in two bands, including blue (B)- and red (R)-band spectra, covering wavelength ranges of, respectively, 4950--5350 {\AA} and 6300--6800 {\AA}, with a spectral resolution of $R = 7500$. 

\subsection{Spectroscopic Observations}

We removed noisy spectra by placing a constraint on the $g$-band signal-to-noise ratios of SNR$_{g} > 20$. The LAMOST database also provides stellar atmospheric parameters ($T_\textrm{eff}$, $\log g$ and [Fe/H]), derived using the LAMOST Stellar Parameter pipeline \citep[LSP3;][]{xiang2015lamost}. For spectra with SNR$_{g} > 30$, the derived stellar parameters have typical uncertainties of $\Delta T_\mathrm{eff} = 90$ K, $\Delta \log g = 0.17$ cm s$^{-2}$ and $\Delta$[Fe/H] = 0.11 dex \citep{li2022estimation}.

\begin{figure}
    \centering
    \includegraphics[width=9cm]{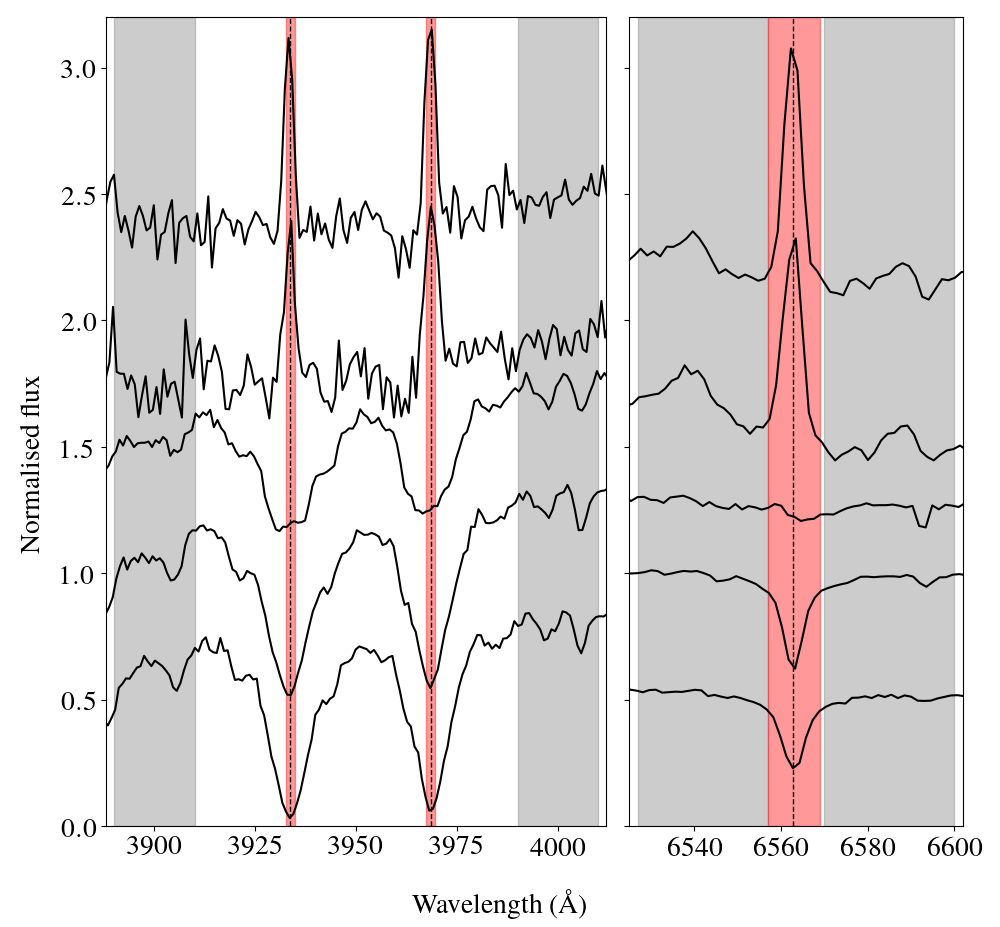}
    \caption{Ca {\sc ii} H and K (left) and H$\alpha$ (right) spectral lines from the LAMOST LRS for the following temperatures, from top to bottom: 3700 K, 4760 K, 5436 K, 5606 K and 5787 K. The red and grey regions are the adopted line-flux bandpasses and continuum levels, respectively.}
    \label{fig:2}
\end{figure}

The LAMOST LRS data contain Ca {\sc ii} H and K and H$\alpha$ chromospheric line emission. The emission profiles of the Ca {\sc ii} H and K lines vary significantly with the level of stellar activity, which is mass-dependent. A sample of LAMOST spectra for different Ca {\sc ii} H and K line profiles is shown in Figure \ref{fig:2}. We can see emission traces in the centres of the absorption troughs in a few of the spectra at the bottom of the left-hand panel of Figure \ref{fig:2}, mainly for G--K-type stars. M-type stars tend to show complete emission profiles (see the top two spectra in the left-hand panel of Figure \ref{fig:2}). A similar profile behaviour is observed for the H$\alpha$ spectra and also for different temperatures (see the right-hand panel of Figure \ref{fig:2}). We have corrected the LAMOST spectra to the rest-frame using the radial velocities derived from the LSP3 pipeline. The resulting LAMOST wavelengths were converted from vacuum to air.

\section{Chromospheric Activity Analysis} \label{sec:section3}

Chromospheric activity indices provide a more accurate estimate of a star's magnetic activity than photospheric activity indices. Photospheric indices tend to become diluted owing to the presence of bright faculae. By contrast, chromospheric indices are measured from line emission, which mainly originates from the stellar chromosphere. Several chromospheric indicators have been used in the literature, including Ca {\sc ii} H and K, Mg {\sc ii} H and K, H$\alpha$ and Na {\sc i} D1 and D2, among others \citep{Hall2008,Saikia2018,silva2021stellar}. In this section, we discuss the techniques used to derive our sample's chromospheric activity indices using spectra containing Ca {\sc ii} H and K and H$\alpha$ line emission.

\subsection{Determination of the Ca {\sc ii} H and K indices}

Chromospheric activity is determined by measuring the flux in the chromospheric Ca {\sc ii} H and K lines and normalising it to the nearby continuum \citep{Saikia2018,silva2021stellar}. We used the ACTIN2\footnote{\url{https://github.com/gomesdasilva/ACTIN2}} \citep{da2018actin} package to estimate the $S$ indices of the Ca {\sc ii} H and K lines, where the $S_\textrm{CaII}$ index is defined as
\begin{equation}
    S_\textrm{CaII} = \frac{F_\textrm{H} + F_\textrm{K}}{R_\textrm{B} + R_\textrm{R}}.
\end{equation}
Here, $F_\mathrm{H}$ and $F_\mathrm{K}$ are the integrated fluxes measured within a triangular bandpass with a full width at half maximum (FWHM) of 1.09 {\AA} for the H and K lines centred at 3968.47 {\AA} and 3933.664 {\AA}, respectively; $R_{\rm B}$ and $R_{\rm R}$ are continuum fluxes measured within rectangular bandpasses of 20 {\AA} centred at 3901.07 {\AA} and 4001.07 {\AA}, respectively \citep{silva2021stellar}. 

\begin{figure}
    \centering
    \includegraphics[width=8.8cm]{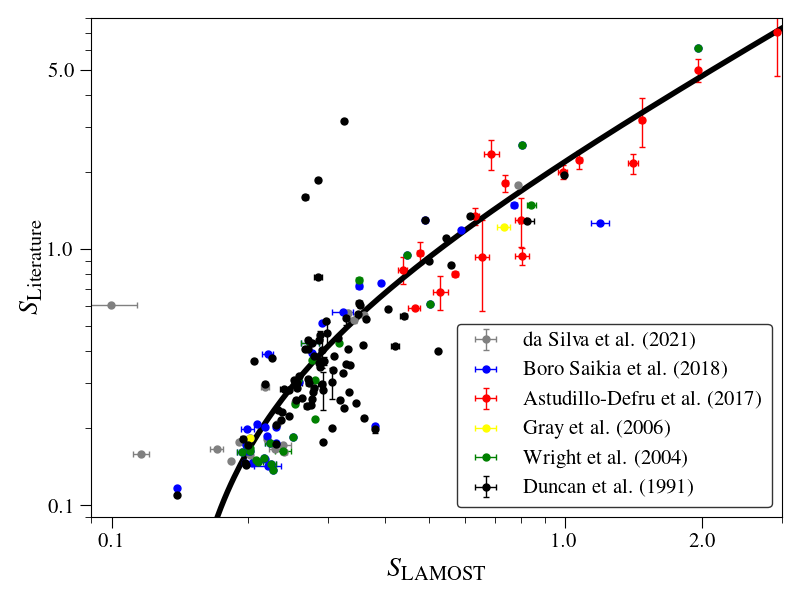}
    \caption{Comparison of $S$ indices from the literature with their LAMOST counterparts, fitted with a linear relation.}
    \label{fig:3}
\end{figure}

Sample spectra with the triangular and continuum bandpasses highlighted for the Ca {\sc ii} H and K lines are shown in Figure \ref{fig:2}. It is clear that the triangular bandpasses with FWHM = 1.09 {\AA} do not cover the full emission-line profiles. Given the low spectral resolution of our data, we suspect that the line emission may have been broadened instrumentally, and therefore it has been spread over an FWHM of 6--8 {\AA}. Hence, by directly using the resulting values we would underestimate the $S$ index since the entire line is not covered. To calibrate the LAMOST $S$ indices against the corresponding Mount Wilson indices ($S_\textrm{MW}$), we found 200, 150, 80, 40, 10 and 6 LAMOST counterparts in, respectively, the \citet{boro2018chromospheric}, \citet{duncan1991ii}, \citet{silva2021stellar}, \citet{Defru2017magnetic}, \citet{wright2004chromospheric} and \citet{gray2006contributions} catalogues. We show the $S$ indices derived from the LAMOST counterparts versus the corresponding literature values in Figure \ref{fig:3}. We calibrated $S_\textrm{LAMOST}$ with respect to $S_\textrm{MW}$ by performing a linear regression on stars in common between LAMOST DR7 and the literature catalogues (see Figure \ref{fig:3}). The expression used for conversion from LAMOST to Mount Wilson $S$ indices is

\begin{equation}
    S_\textrm{MW} = 2.55(\pm 0.15) (S_\textrm{LAMOST}) - 0.346(\pm 0.05).
\end{equation}

We observe a dispersion in the $S$ index from the linear relation between the literature and LAMOST-derived values. This could have been caused by the fact that line-profile broadening is not linear, owing to the low resolution of our LAMOST LRS data, or perhaps because LAMOST may have measured the activity at different times (in some cases, the difference in observation epoch is up to two decades).

The $S$ index is colour-dependent, and hence we converted it to a chromospheric flux ratio, $\log R'_\textrm{HK}$. The $R'_\textrm{HK}$ index can be calculated from $S_\textrm{MW}$ using the \citet{noyes1984rotation} expression,
\begin{equation}
    R'_\textrm{HK} = R_\textrm{HK} - R_\textrm{phot};  \\
    R_\textrm{HK} = 1.34\cdot 10^{-4} \cdot C_\textrm{cf}\cdot S_\textrm{MW},
\end{equation}
where $R_\textrm{HK}$ is the index corrected to relate to the bolometric flux, $C_\textrm{cf}$ is the bolometric correction, and $R_\textrm{phot}$ is the photospheric contribution. We used the photospheric contribution as a function of $(B-V)$ colour as derived by \citet{noyes1984rotation}. \citet{mascareno2015rotation} and \citet{mascareno2016magnetic} derived the bolometric correction factor, $C_\textrm{cf}$, and extended its applicability to colours as red as $(B-V)=1.9$ mag, which includes M dwarfs. Since our catalogue also includes M-type stars, we used the \citet{mascareno2015rotation,mascareno2016magnetic}-derived correction factor.

\subsection{Determination of H\texorpdfstring{$\alpha$}{h} indices}

H$\alpha$ emission arises owing to collisional excitation in the densest region of the stellar chromosphere. Hence, it is the strongest and most widely used magnetic activity indicator \citep{fang2016stellar,fang2018stellar}. We used Fang et al.'s method to calculate our sample's excess fractional luminosities for the H$\alpha$ line, $\log R'_{\textrm{H}\alpha}$. Since the H$\alpha$ line is present in both the low- and medium-resolution data, we estimated $\log R'_{\textrm{H}\alpha}$ for both data sets.

\subsubsection{Equivalent Widths}

\begin{figure}
    \centering
    \includegraphics[width=8.75cm]{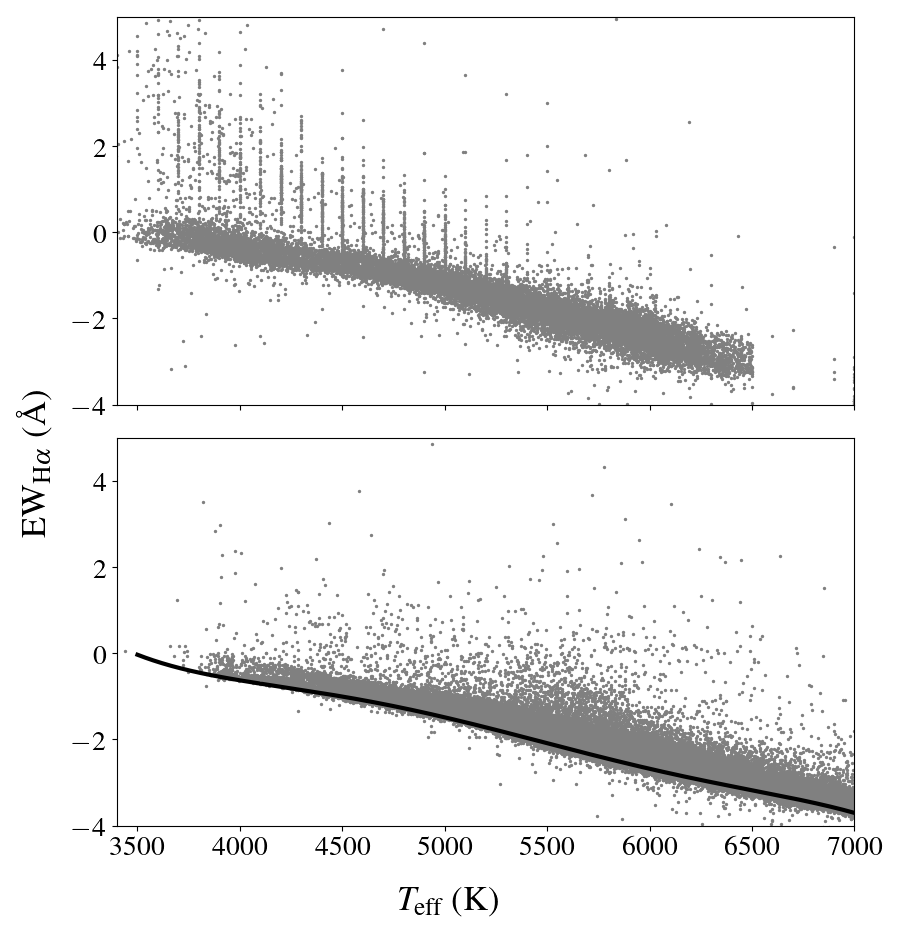}
    \caption{H$\alpha$ EW versus effective temperature (top) for our target sample, derived from the LAMOST LRS spectra. EWs derived for 300,000 LAMOST LRS spectra (right), where the solid line indicates the 10 per cent quantile representing the basal EW for low-mass stars.}
    \label{fig:4}
\end{figure}

Equivalent widths (EWs) were measured from the radial velocity-corrected LAMOST spectra. H$\alpha$-line EWs were estimated using the expressions of \citet{fang2016stellar,fang2018stellar}. Measurements of the H$\alpha$ line were centred at 6563 {\AA} across a 12 {\AA} bandpass, whereas the continuum flux was estimated by interpolating the fluxes between the continuum ranges at 6527--6557 {\AA} and 6570--6600 {\AA}. We increased the continuum ranges as discussed in detail in Appendix \ref{sec:section7}. A sample of H$\alpha$ spectra with their bandpasses and continua is shown in Figure \ref{fig:2} for different effective temperatures. The derived EW$_{\textrm{H}\alpha}$ values are plotted against effective temperature in Figure \ref{fig:4}. The data points appearing as vertical bars in the top panel of Figure \ref{fig:4} are taken from our catalogue of BY Draconis variables \citep{chahal2022statistics}. They were derived based on SED fitting using the VOSA tool, with a step size of 100K. To measure the excess EWs, EW$'_{\textrm{H}\alpha}$, basal lines are required. Basal-line measurements correspond to the reference EW for magnetically inactive stars. To estimate the basal value, we selected sources satisfying 3300 K $\leq T_\textrm{eff} \leq$ 7000 K, $\log g \geq 4.0$ dex and SNR$_{g} \geq 100$ from LAMOST DR7, resulting in approximately 300,000 objects. We calculated their EWs (see the bottom panel of Figure \ref{fig:4}) and binned them using a bin width of 50 K. In each bin, the 10 per cent quantiles were calculated for EW$_{\textrm{H}\alpha}$ \citep{zhang2019stellar}. Following \citet{zhang2019stellar}, a fifth-order polynomial was fitted to the quantiles, shown as the solid line in Figure \ref{fig:4}. This solid line was treated as the minimum or basal value pertaining to the H$\alpha$ chromospheric emission. EW$'_{\textrm{H}\alpha}$ values were estimated by subtracting the basal value for the same effective temperature, EW$_{\textrm{H}\alpha,\textrm{basal}}$, i.e.,

\begin{equation}
    \textrm{EW}'_{\textrm{H}\alpha} = \textrm{EW}_{\textrm{H}\alpha} - \textrm{EW}_{\textrm{H}\alpha,\textrm{basal}}.
\end{equation}

\subsubsection{Excess fractional luminosities}

Conversion to fractional luminosity is required to remove the dependence of EW$'_\mathrm{H\alpha}$ on the surrounding continuum level and to compare activity levels across different spectral types. The excess fractional luminosity is defined as the ratio of the H$\alpha$ luminosity to the bolometric luminosity. It can be derived from the excess EW using a distance-independent value, $\chi$, i.e., the ratio between the continuum flux near the line of interest ($F(\lambda_{c})$) and the bolometric flux \citep{west2004spectroscopic,fang2016stellar,fang2018stellar}:

\begin{equation}
    R'_{\textrm{H}\alpha} \equiv \chi_{\textrm{H}\alpha}\cdot \textrm{EW}'_{\textrm{H}\alpha} = \frac{F(\lambda_{c})}{\sigma T^{4}_\textrm{eff}}\cdot \textrm{EW}'_{\textrm{H}\alpha},
\end{equation}
where $\sigma = 5.6704\times 10^{-5}$ erg$^{-1}$ cm$^{-2}$ K$^{-4}$ is the Stefan--Boltzmann constant. The surface continuum flux, $F(\lambda_{c})$, for the H$\alpha$ line, was estimated by measuring the continuum flux from synthetic spectra generated using the PHOENIX stellar atmosphere models \citep{husser2013new}. We used the {\tt pysynphot} python package to generate synthetic spectra for solar metallicity and solar surface gravity. We compared the H$\alpha$ indices derived from both the low- and medium-resolution spectra and found that both indices are similar.

\begin{table*}
\caption{Compiled table of sources taken from \citet{Mcquillan2014} (Kepler), \citet{Reinhold2020} (K2) and \citet{chahal2022statistics} (ZTF) with their stellar parameters. We have also presented the chromospheric activity indices derived for Ca {\sc ii} H $\&$ K and H$\alpha$ line emissions using the LAMOST spectra.}
\centering
\resizebox{\textwidth}{!}{
\begin{tabular}{cccccccccccccccc}
\hline
ID \\ ZTF \ \ \ \ \ EPIC \ \ \ \ \ KIC &
  \begin{tabular}[c]{@{}c@{}}R.A. (J2000)\\ ($^{\degree}$)\end{tabular} &
  \begin{tabular}[c]{@{}c@{}}Dec. (J2000)\\ ($^{\degree}$)\end{tabular} &
  \begin{tabular}[c]{@{}c@{}} LAMOST \\ OBSID \end{tabular} &
  \begin{tabular}[c]{@{}c@{}}$T_{\rm eff}$\\ (K) \end{tabular} &
  \begin{tabular}[c]{@{}c@{}}BP-RP\\ (mag) \end{tabular} &
  \begin{tabular}[c]{@{}c@{}}$P_\mathrm{rot}$\\ (d) \end{tabular} &
  \begin{tabular}[c]{@{}c@{}}RV\\ (km s$^{-1}$)\end{tabular} &
  \begin{tabular}[c]{@{}c@{}} LRS \\ EW \end{tabular} &
  \begin{tabular}[c]{@{}c@{}} LRS \\ log $R'_{\rm H\alpha}$ \end{tabular} &
  \begin{tabular}[c]{@{}c@{}} MRS \\ EW \end{tabular} &
  \begin{tabular}[c]{@{}c@{}} MRS \\ log $R'_{\rm H\alpha}$ \end{tabular} &
  \begin{tabular}[c]{@{}c@{}}$S_\mathrm{MW}$\\ \end{tabular} &
  \begin{tabular}[c]{@{}c@{}}$S_\mathrm{err}$\\ \end{tabular} &
  \begin{tabular}[c]{@{}c@{}}log $R'_\mathrm{HK}$\\ \end{tabular} \\ \hline
ZTFJ000000.13+620605.8 & 102.13442 & 21.13442 & 95608182 & 3856 & 2.655 & 17.95 & 34.75  & 1.79  & -3.85 & 2.587  & -3.725  & 11.148 & 0.0686  & -4.735 \\
ZTFJ000000.51+583238.7 & 92.00221 & 23.91978 & 617815053 & 6029 & 0.958 & 2.57 & -7.425  & -1.954 & -4.079 & -1.478 & -3.867 & 0.455 & 0.0106 & -4.422 \\
ZTFJ000002.20+480720.8 & 95.71082 & 20.44823 & 437312203 & 3592 & 2.2   & 15.96 & 23.43    & 0.038 & -4.978 & -0.0447 & -5.178 & 1.146 & 0.0148  & -5.64 \\
ZTFJ000003.23+543605.4 & 99.45448 & 18.19605 & 688114016 & 5990 & 1.008 & 2.41 & 7.9    & -1.245 & -3.8 & -1.059 & -3.75 & 0.755 & 0.0041  & -4.23 \\
ZTFJ000003.76+532917.1 & 93.56008 & 25.61492 & 486311029 & 4873 & 1.854 & 15.2 & 9.6   & -0.486 & -4.032 & -0.14 & -3.88 & 1.31  & 0.012  & -5.154 \\
...                    & ...     & ...      & ... & ...  & ... & ...  & ...    & ...   & ...   & ...   & ...  & ...   & ...   & ...  \\
...                    & ...     & ...      & ... & ...  & ... & ...  & ...    & ...   & ...   & ...   & ...  & ...   & ...   & ...  \\ \hline

\end{tabular}}

Only a portion of the table is shown here. The full catalogue is available in the online journal.

\label{table1}
\end{table*}

\section{Investigating Chromospheric Activity} \label{sec:section4}

Stars are born with a range of rotation periods \citep{angus2015calibrating}. Rapidly rotating stars lose angular momentum quickly in comparison with slow rotators and converge onto a well-defined slowly rotating sequence. The time-scale of convergence is expected to be around the age of the Praesepe cluster for solar-mass stars, i.e., approximately 670 Myr \citep{angus2015calibrating}. After this age, stellar rotation periods can be considered independent of their initial values. The databases published by \citet{Mcquillan2014} and \citet{Reinhold2020} predominantly contain slowly rotating stars, i.e., stars that have already converged onto the slowly rotating sequence. Both catalogues contain a wide range of spectral types, from F to M. These databases exhibit a well-known period gap \citep[e.g.,][]{Mcquillan2014,curtis2019temporary,curtis2020stalled,Reinhold2020,Gordon2021}. The origin of this gap is still subject to discussion, however. 

Our ZTF catalogue of BY Dra variables \citep{chahal2022statistics} also shows tentative evidence of a period gap. BY Dra are main-sequence FGKM-type stars that show modulation in their luminosities owing to the rotation of starspots. Our BY Dra catalogue contains mainly rapidly rotating sources since 90 per cent of our objects have $P_\mathrm{rot} \leq$ 10 days. Therefore our BY Dra catalogue complements the \citet{Mcquillan2014} and \citet{Reinhold2020} catalogues. Therefore, we have combined all three catalogues and derived the Ca {\sc ii} H and K and H$\alpha$ chromospheric activity indices for all of their LAMOST counterparts. In this section, we present the dependence of the activity indices on different stellar parameters. We also discuss the trends we have found in the period--colour diagram. A sample of the final table containing source coordinates, de-reddened {\sl Gaia} colours, stellar parameters, and the derived activity indices is shown in Table \ref{table1}. The entire table is available electronically in the online journal.

\subsection{Activity Dependence on Stellar Parameters}

\begin{figure}
    \centering
    \includegraphics[width=8.75cm]{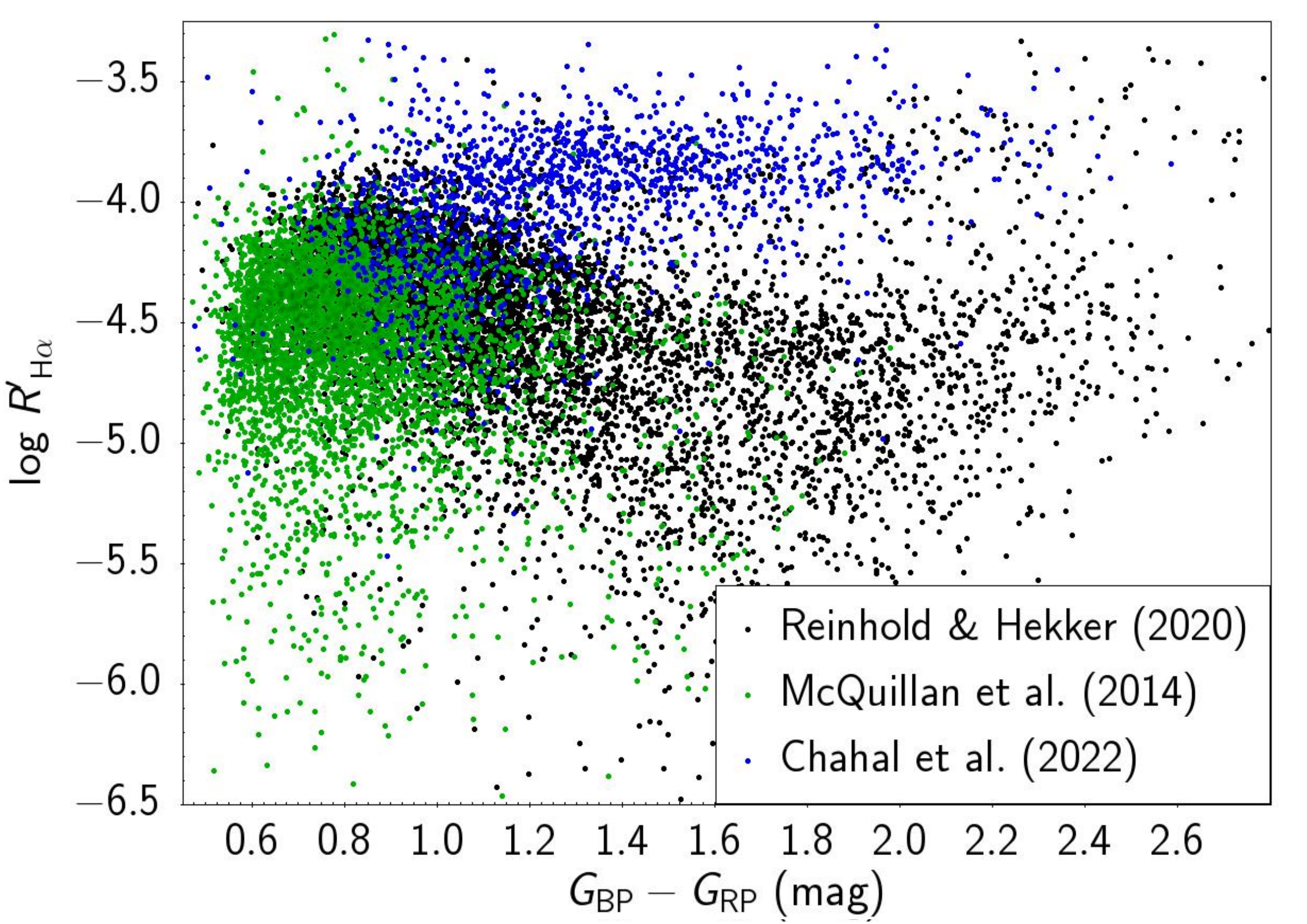}
    \caption{$\log R'_\mathrm{H\alpha}$ versus {\sl Gaia} colours (de-reddened), showing data from three different catalogues covering different ranges of effective temperature and activity indices.}
    \label{fig:5}
\end{figure}

We derived chromospheric activity indices from our sample's Ca {\sc ii} H and K and H$\alpha$ line emission lines (see section \ref{sec:section3}). An example distribution of the H$\alpha$ indices ($R'_{\rm H\alpha}$) is shown in Figure \ref{fig:5}. Note that the Ca {\sc ii} H and K indices are affected by uncertainties owing to broadened line emission associated with our use of low-resolution spectra. However, we have compared the trends in the derived log $R'_{\rm HK}$ indices with those obtained by \citet{noyes1984rotation,mamajek2008improved,boro2018chromospheric} and found them to exhibit similar behaviour.
We discuss the behaviour of the Ca {\sc ii} H and K indices in the Appendix. In Figure \ref{fig:5}, data from the different surveys are plotted as a function of {\sl Gaia} colour. The three data sets cover different ranges in mass, rotation period, and chromospheric activity. \citet{Mcquillan2014} mainly cover chromospheric activity for F--G- and a few K-type stars. They exhibit a broad range of chromospheric activity. The \citet{Reinhold2020} data set covers G--K--M-type stars. These stars exhibit a clear decrease in activity towards lower masses. Our ZTF sample \citep{chahal2022statistics}, on the other hand, which contains mostly rapidly rotating stars, shows increased activity for lower masses. The activity indices for K--M-type stars in the ZTF \citep{chahal2022statistics} and K2 \citep{Reinhold2020} surveys show a different type of behaviour, likely owing to the difference in their rotation rates. 

\begin{figure*}
    \centering
    \includegraphics[width=18.25cm]{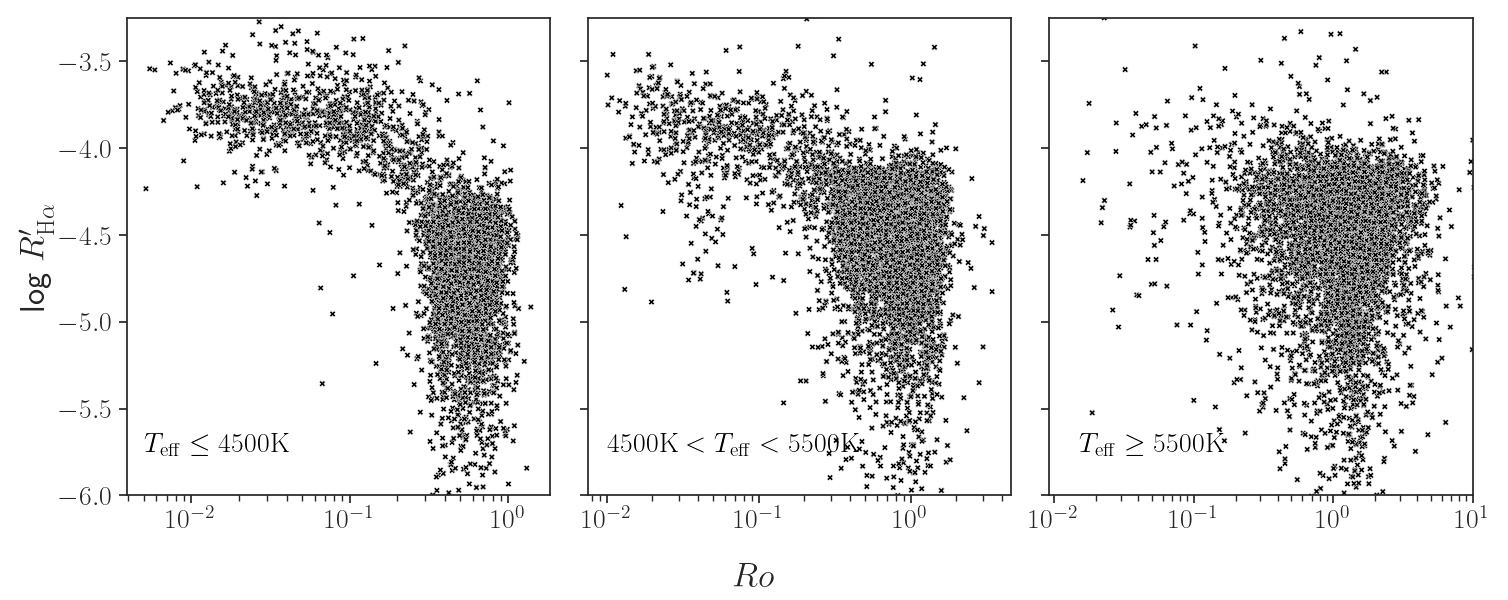}
    \caption{Chromospheric Activity indices, $\log R'_{\textrm{H}\alpha}$ index, plotted as a function of Rossby number ($Ro \equiv P_\mathrm{rot}/\tau_\mathrm{conv}$) for different temperature ranges in three panels.}
    \label{fig:6}
\end{figure*}

To understand the activity dependence on stellar parameters, we have plotted the corresponding Rossby number ($Ro$) in Figure \ref{fig:6}. The Rossby number is defined as the ratio of the stellar rotation period to the convective turnover time-scale, $Ro \equiv P_\mathrm{rot}/\tau_\mathrm{conv}$. It reflects the interplay between rotation and convection, which is, in turn, responsible for driving a star's dynamo action. The convective turnover time is estimated from an analytical equation derived from a parameterised fit, as a function of effective temperature, obtained by \citet{Cranmer2011}. To understand the behaviour of chromospheric activity as a function of the Rossby number, we show the distribution of both parameters in Figure \ref{fig:6}. We have segregated our data set based on effective temperature (see Figure \ref{fig:6}). The chromospheric activity versus Rossby number in Figure \ref{fig:6} clearly shows the activity behaviour similar to as observed with photospheric activity \citep[see][]{see2021photometric} and X-ray Luminosity \citep[see][]{reiners2014generalized}. In the left panel, we do not observe the slope in activity with Rossby number for stars with $Ro \geq 0.3$, and this trend goes to higher Rossby numbers with higher temperatures.

\subsection{Comparative Analysis of Star Clusters}

To study the evolution of magnetic activity and, hence, to quantify the level of spin-down or magnetic braking, it is important to have accurate stellar age estimates. Unfortunately, the estimation of accurate ages, especially for main-sequence stars, is very difficult. Over the past two decades, catalogues of star clusters have increased significantly in volume. All members in a given star cluster are approximately coeval. Hence, age estimates of star clusters derived from isochrone fitting are quite precise. Recently, \citet{cantat2020clusters} compiled a list of members of 1481 known star clusters using {\sl Gaia} DR2 data. We cross-matched the star clusters' membership catalogue (435,833 stars) with LAMOST DR7. We thus obtained LAMOST spectra for ~3800 cluster members (see Figure \ref{fig:7} in the Appendix). We obtained spectra for star cluster members covering a wide range of ages, which will assist us in understanding the evolution of stellar chromospheric activity (see Figure \ref{fig:10}). We subsequently derived their H$\alpha$ chromospheric activity indices ($R'_\mathrm{H\alpha}$), as discussed in Section \ref{sec:section3}. The behaviour of the activity--age relation for these cluster members, for different spectral types, is discussed in the next section. 

\subsection{Activity--Rotation--Colour diagram}

\begin{figure*}
    \centering
    \includegraphics[width=18cm]{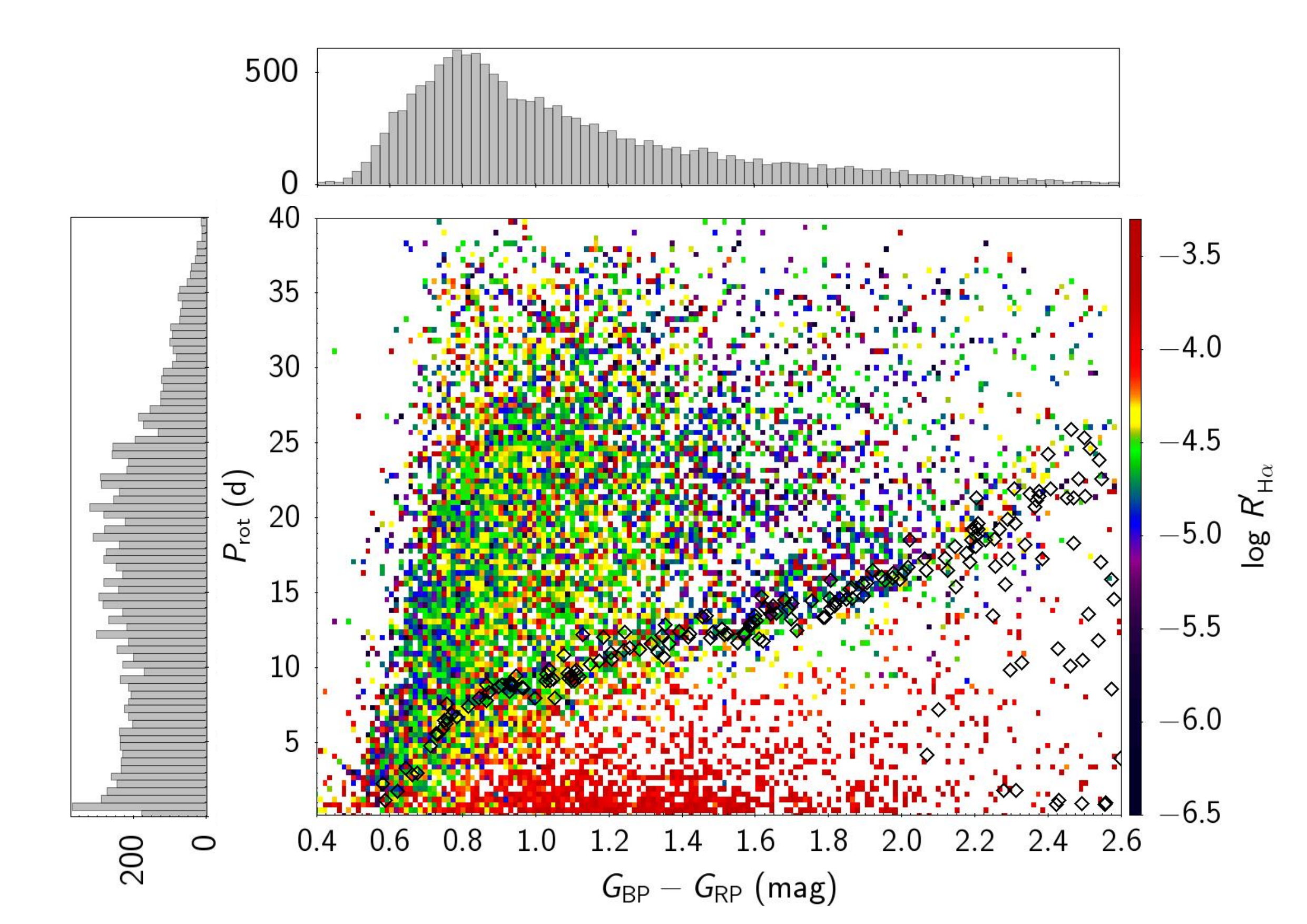} 
    \hspace*{2.8cm} 
    \includegraphics[width=14.6cm]{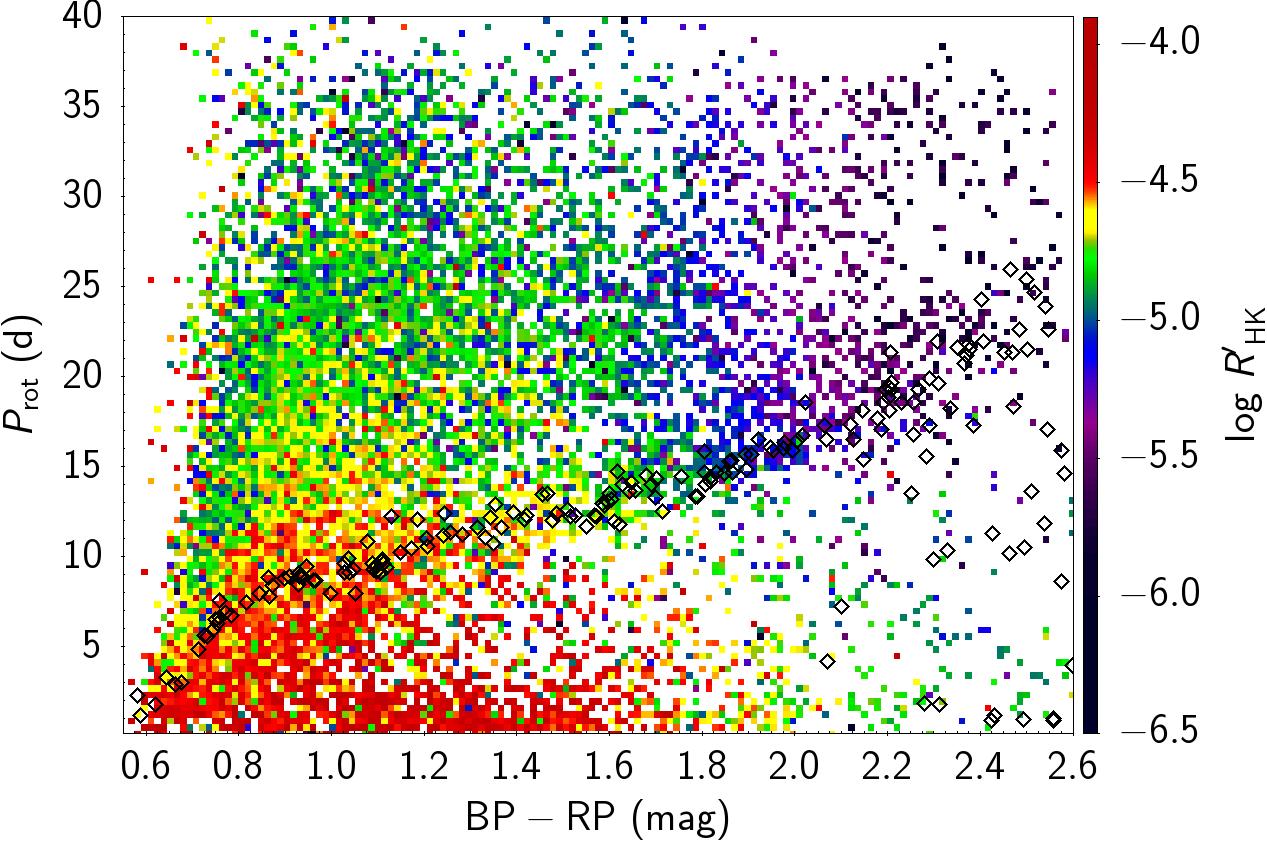}
    \caption{Rotation period versus {\sl Gaia} colour (de-reddened), with (top) H$\alpha$ and (bottom) Ca {\sc ii} H and K activity indices colour-coded. The Praesepe cluster ($\sim$670 Myr) is shown as black data points. Histograms of the rotation periods and {\sl Gaia} colours are shown in the top panel.}
    \label{fig:8}
\end{figure*}

To understand the evolution of stellar rotation and magnetic activity, we plotted the stellar rotation periods as a function of colour in Figure \ref{fig:8}. The distributions of the chromospheric activity indices derived from the Ca {\sc ii} H and K and H$\alpha$ line emission are colour-coded. We have binned the data in Figure \ref{fig:8} to emphasise trends in the activity indices. Measurements of rotation rates in star clusters have shown that clusters older than the Praesepe cluster ($\sim$670 Myr) exhibit spin-down stallation \citep[e.g.,][]{agueros2018new,curtis2019temporary,curtis2020stalled}. Hence, to understand the level of spin-down stallation among our sample stars in relation to the locus of the Praesepe cluster, we have also included Praesepe cluster data points in Figure \ref{fig:8}. The Praesepe cluster data overlap with the region where we see evident changes in the activity indices for different stellar spectral types. Hence, we will refer to stars located below the Praesepe cluster locus as the rapidly rotating region and those above that locus as the slowly rotating region. We discern high activity indices for stars located in the rapidly rotating region. We also observe a clear reduction in activity as we transition from the rapidly to the slowly rotating regime. The reduction in activity is greater for low-mass stars, especially for M-type stars.

Along the slowly rotating sequence, we observe that the activity indices are lower for low-mass stars, i.e., partially convective K--M-type stars exhibit lower activity levels than F--G-type stars, which have thinner convective envelopes. We note that the observed trends in activity are similar for both activity indices, $\log R'_\mathrm{H\alpha}$ and $\log R'_\mathrm{HK}$, which are derived from, respectively, the H$\alpha$ and the Ca {\sc ii} H and K lines (see Figure \ref{fig:8}). We suspect that the low level of activity in the slowly rotating region, especially for K--M-type stars, is so low because these stars are in a spin-down stallation stage \citep{curtis2020stalled}. We will discuss this mechanism in more detail in the next section. Another point to note is that the activity indices are also rather low around the location of the period gap (location is more prominent in Figure \ref{fig:13}). Hence, the period gap seems to be related to spin-down stallation in some way, which will also be discussed in the next section.

\section{Discussion} \label{sec:section5}

Solar magnetic activity originates from the solar dynamo process. The most widely accepted dynamo theory is that of a two-layer interface dynamo \citet{parker1993solar}, where toroidal and poloidal fields are produced in adjacent regions, coupled by diffusion. Strong toroidal fields are produced by strong differential rotation, which is present at the tachocline, i.e., the transition between the radiative core and the convective envelope. The poloidal field is produced by the so-called `$\alpha$ effect' operating on the toroidal field, which diffuses into the lower convection zone \citep{montesinos2001new}. Helioseismic measurements \citet{charbonneau1999helioseismic} support the presence of strong radial shear and differential rotation between the convective zone and the nearly rigidly rotating radiative interior. However, it is still unclear whether or not similar dynamo processes occur in stars of different spectral types.

Magnetic activity generated through the stellar dynamo process leads to mass loss through flares, coronal mass ejections, and other mechanisms. Owing to this mass loss, stars must spin down to conserve their angular momentum, a process referred to as magnetic braking. As stars evolve, several studies \citep{Mcquillan2014,Reinhold2020,Gordon2021} have shown a dearth of sources around a fairly well-defined locus in the period--colour diagram. This period gap is suspected to arise because of spin-down stallation. Recently, \citet{agueros2018new} and \citet{curtis2019temporary,curtis2020stalled} argued that stars halt their spinning down when they reach an age similar to that of the Praesepe cluster ($\sim$670 Myr). \citet{agueros2018new} found that the 1.4 Gyr-old star cluster NGC 752 overlaps with the Praesepe sequence. \citet{curtis2019temporary} showed that the 1 Gyr-old star cluster NGC 6811 also merges seamlessly with the Praesepe cluster sequence. 

These studies have shown that stars cease spinning down once they merge onto the slowly rotating sequence, and they remain stalled for extended periods of time before resuming their spin-down. Based on the loci where the Praesepe, NGC 6811, and NGC 752 sequences overlap, the duration of this temporary period of stalled spin-down appears to increase towards lower stellar masses. \citet{curtis2020stalled} extended their study to older star clusters, including Ruprecht 147 (2.7 Gyr) and NGC 6819 (2.5 Gyr). These authors found that comparatively high-mass stars resume spinning down, while M-type stars still merge towards the Praesepe cluster sequence. They concluded that spin-down stallation is a temporary phase and that its duration is mass-dependent. The period gap arises because of a temporarily stalled spin-down at the location of the period gap, followed by a rapid spin-down. This creates a dearth of sources around the location of the period gap. Recently \citet{lu2022bridging}, found that the period gap closes at the fully convective limit. This supports the notion that the gap may have been created because of core--envelope coupling, which is the most frequently discussed hypothesis underlying spin-down stallation. In this section, we will discuss the evidence of the spin-down stallation mechanism with respect to chromospheric activity and its possible link to the core--envelope coupling. 

\subsection{Evidence of Spin-down Stallation}

\begin{figure*}
    \begin{multicols}{2}
    \includegraphics[width=1.05\linewidth]{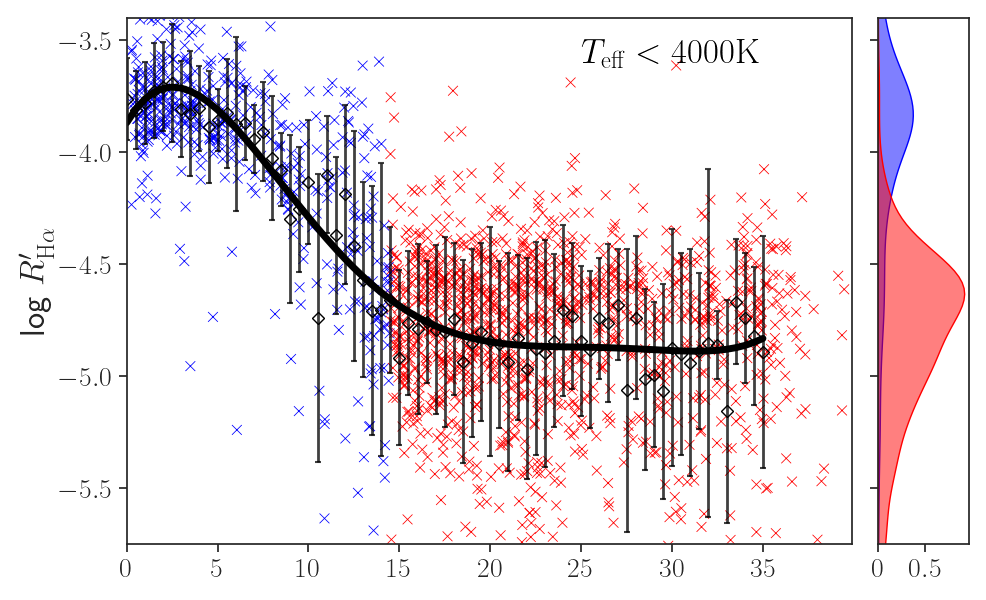}\par 
    \includegraphics[width=1\linewidth]{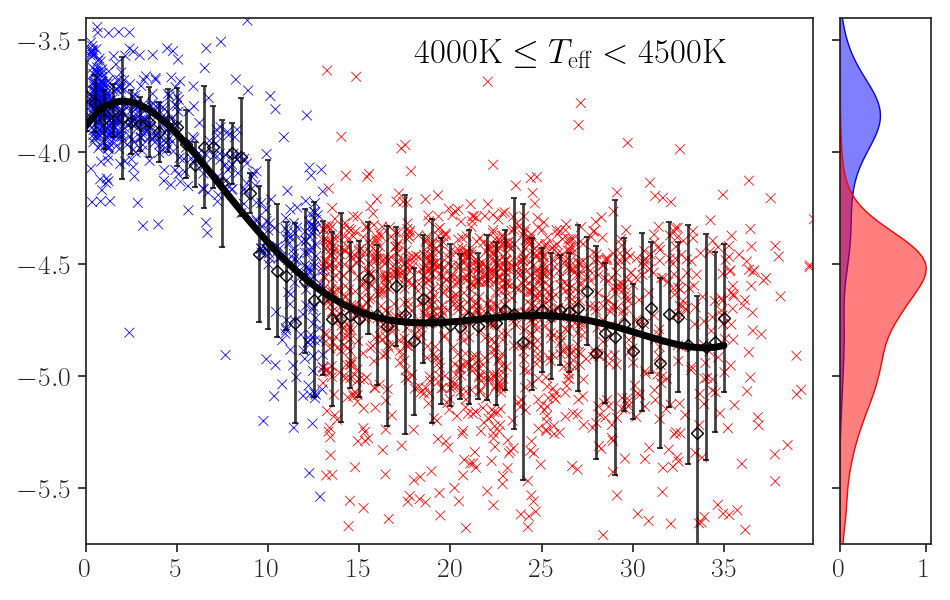}\par 
    \end{multicols}
    \begin{multicols}{2}
    \includegraphics[width=1.05\linewidth]{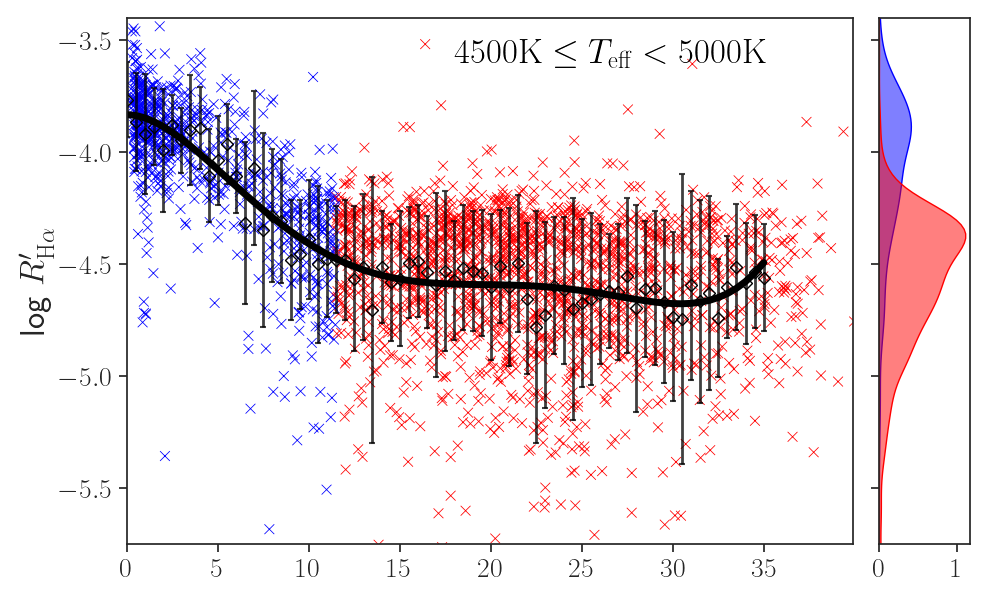}\par 
    \includegraphics[width=1\linewidth]{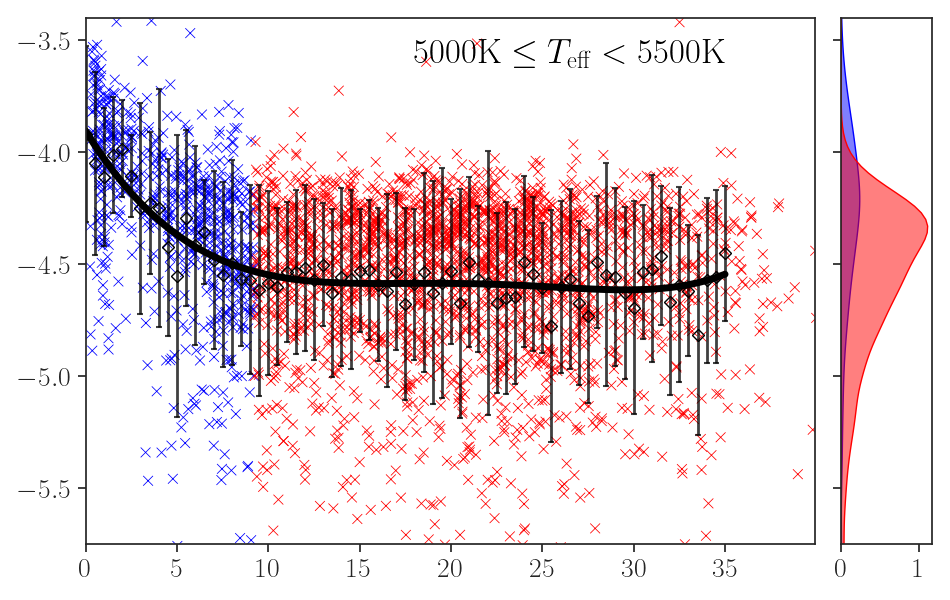}\par 
    \end{multicols}
    \begin{multicols}{2}
    \includegraphics[width=1.05\linewidth]{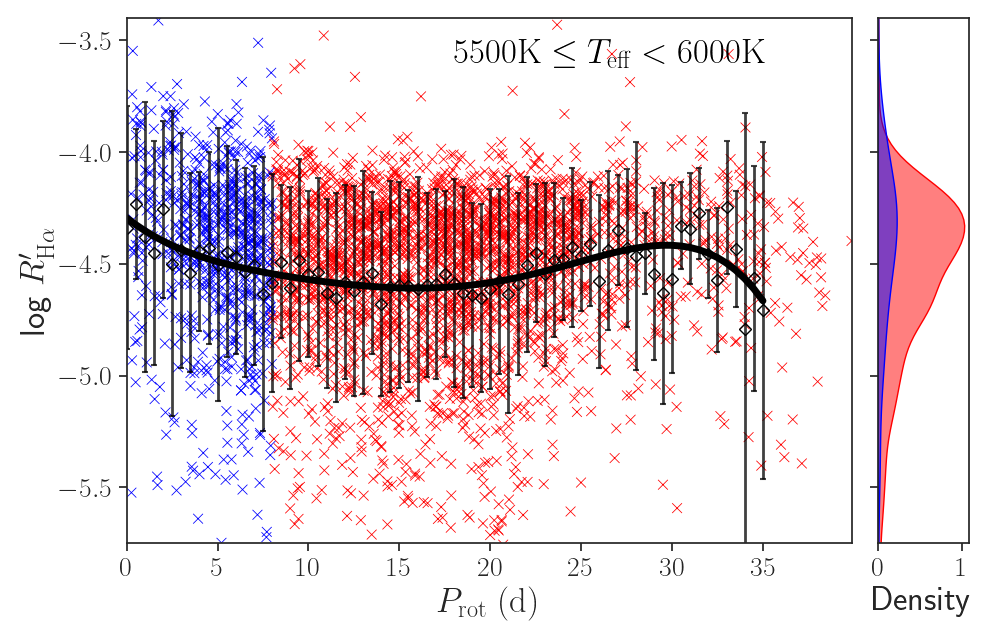}\par 
    \includegraphics[width=1\linewidth]{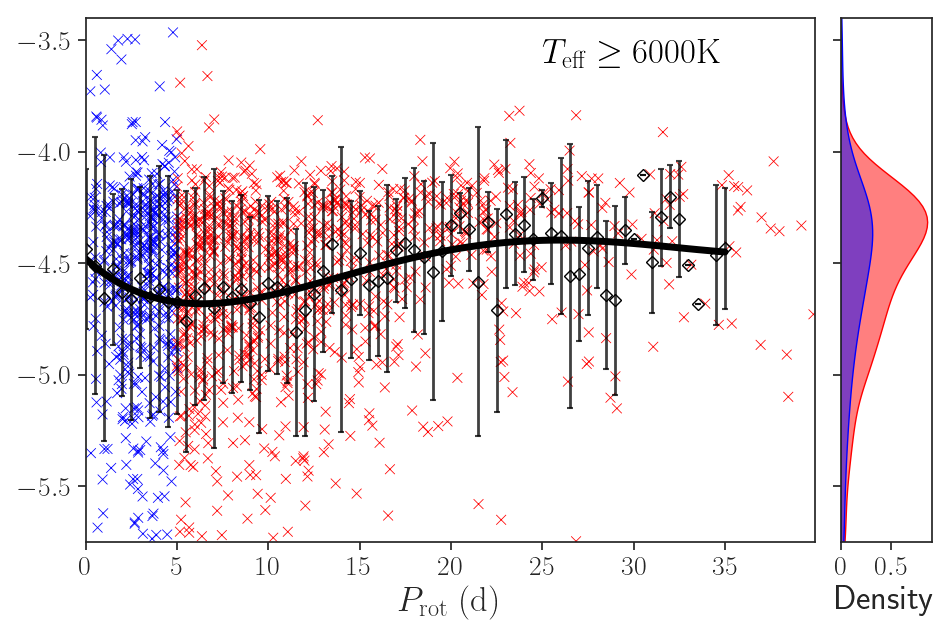}\par 
    \end{multicols}
        \caption{Chromospheric activity indices, $\log R'_\mathrm{H\alpha}$, as a function of rotation period for different effective temperature ranges. Blue data points represent the rapidly rotating sequence, within their respective temperature ranges. These stars are younger than the Praesepe cluster ($\sim$670 Myr). The red data points represent a slowly rotating sequence, including stars that are older than the Praesepe cluster. The data points have been binned using bin widths of 0.5 days, and their respective means and standard deviations are also shown. The solid line represents the best-fitting fifth-order polynomial to the mean values. The histograms on the right of each panel show the density distributions of the activity indices for the blue and red data points.}
    \label{fig:9}
\end{figure*}

Based on previous studies of the locations of different star clusters in the rotation--colour diagram \citep{agueros2018new,curtis2019temporary,curtis2020stalled,david2022further}, spin-down stallation appears to occur around the age of the Praesepe cluster. We have tried to understand this phenomenon based on the derived chromospheric activity indices. We show the H$\alpha$ activity indices versus the stellar rotation periods for different effective temperature ranges in Figure \ref{fig:9}. To understand the onset of the spin-down stallation stage with respect to the age of the Praesepe cluster, we coloured rapidly rotating stars compared with Praesepe cluster members blue, and slowly rotating stars red, within the respective temperature ranges. We note some trends in the data for different temperatures in Figure \ref{fig:9}. To make these trends more prominent, we binned the data based on the rotation periods and estimated their mean values (shown as black open data points with their respective standard deviations in Figure \ref{fig:9}). Next, to visualise the behaviour of the mean activity, we fitted the data with polynomials of different orders. We found that a fifth-order polynomial yielded the optimal fit. We also plotted the Ca {\sc ii} H and K indices and rotation periods for different temperature ranges in Figure \ref{fig:12} (see Appendix \ref{sec:section6}).

Based on Figure \ref{fig:9}, it is evident that the rapidly rotating stars (see the blue data points) exhibit a steep decrease in activity. The Praesepe cluster is located in the region where the steep slope of activity gradually decreases. The activity saturates for slowly rotating stars (see the red data points). Along the rapidly rotating sequence, the steep decrease in activity is reduced towards higher temperatures and converges towards constant activity. Along the slowly rotating sequence, we observe an increase in saturated activity with a commensurate increase in temperature. Based on the activity histograms shown in Figure \ref{fig:9}, we clearly see that both the slowly and rapidly rotating sequences merge with each other, i.e., the mean activity along the rapidly rotating sequence decreases, whereas, along the slowly rotating sequence, it increases with increasing temperature. 

Figure \ref{fig:9} also contains indices derived from our medium-resolution spectra, wherever available. We have found that indices obtained from medium-resolution spectra exhibit similar activity trends, as depicted in Figure \ref{fig:9}. We also see a similar decrease in the Ca {\sc ii} H and K indices for different temperature ranges (see Figure \ref{fig:12}). However, instead of saturated low activity, we observe a reduction in activity for the Ca {\sc ii} H and K indices along the slowly rotating sequence (see the red data points in Figure \ref{fig:12}). The observed decrease might be owing to comparatively higher uncertainties (discussed in Section \ref{sec:section3}), which are difficult to quantify. We have also found similar activity trends across different temperature ranges in photospheric activity indices taken from \citet{Mcquillan2014,Reinhold2020}. \citet{Reinhold2020} have also discussed the decreased photospheric activity in detail and suggested the period gap arises because of the decreased activity.

We suspect that this steep decrease in activity and spin-down stallation are both caused by core--envelope coupling, the process suggested by \citet{curtis2020stalled}. We suggest that core--envelope coupling commenced earlier than the age of the Praesepe cluster, i.e., before an age of 670 Myr, somewhere along the rapidly rotating sequence where the activity slope starts to decline. The slope of activity changes around an age of $\sim$700--1000 Myr, where the spin-down stallation has been observed.
We also suggest that spin-down stallation may be effective for low-temperature stars, especially for partially convective K--M-type stars, since they clearly show a steep decline and then a constant level of low activity. Meanwhile, for relatively high-mass stars, i.e., for F--G-type stars, the saturated activity level is relatively higher, which indicates that the core--envelope coupling and hence the spin-down stallation might not be very effective in them.

\subsection{Activity--Age relation in Star Clusters}

Rotation and the convective envelope are the major drivers of stellar dynamo action, and hence of magnetic activity. Stars lose angular momentum through magnetised stellar winds, hence leading to spin-down. As spin-down happens, the regeneration of the magnetic field also weakens with time. Therefore, we expect the magnetic activity to decrease as a star evolves. Several studies have tried to calibrate the activity--age relation so as to use it as a standard for age estimation \citep{barnes2007ages,mamajek2008improved,pace2013chromospheric}. This dependence on mass and evidence of spin-down stallation suggest that the relationship cannot be simply fitted with a single power-law function. \citet{pace2013chromospheric} studied the activity--age relation. They found that chromospheric activity exhibits a steep decrease, and it does not decay any further after an age of 2 Gyr. We see similar behaviour in our data (see Figure \ref{fig:9}).

We have derived chromospheric activity indices for our star cluster members. The activity--age distribution for different effective temperature ranges is shown in Figure \ref{fig:10}. The means and standard deviations for the cluster members are shown in Figure \ref{fig:10}. To capture the distribution's main large-scale features, the mean values were fitted with different polynomial orders; we found a good fit for a fourth-order polynomial. The Sun's locus is also shown in Figure \ref{fig:10}, for comparison (red symbol). Solar spectra were taken from the HARP\footnote{https://www.eso.org/sci/facilities/lasilla/instruments/harps/inst/monitoring/sun.html} Solar Spectra Collection. We observe a decrease in activity index with increasing age, but the decay slope is different for different stellar masses. K--M-type stars exhibit high activity around the ages of 5--10 Myr, followed by a steep decrease in activity until 1--2 Gyr when the activity level is even lower than that of the Sun. For F--G-type stars, the activity decrease is not very steep and reaches an approximately saturated level for stars older than 100 Myr. This trend in activity closely follows the Sun's activity. 

{\begin{figure*}
    \includegraphics[width=18cm]{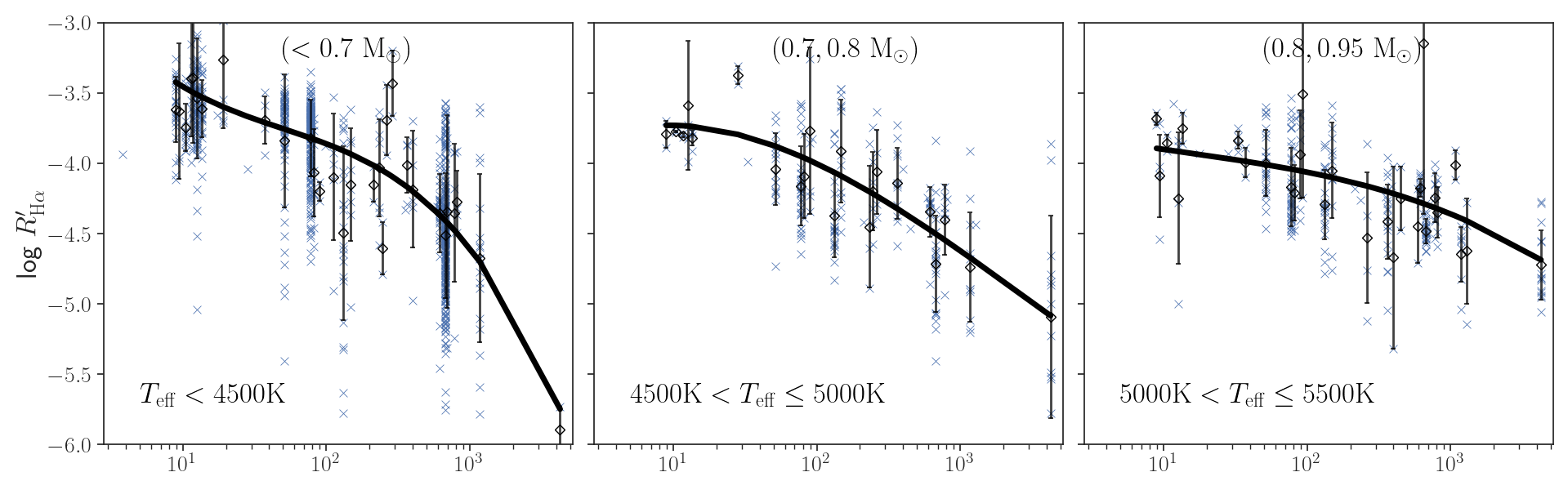}
    \includegraphics[width=18cm]{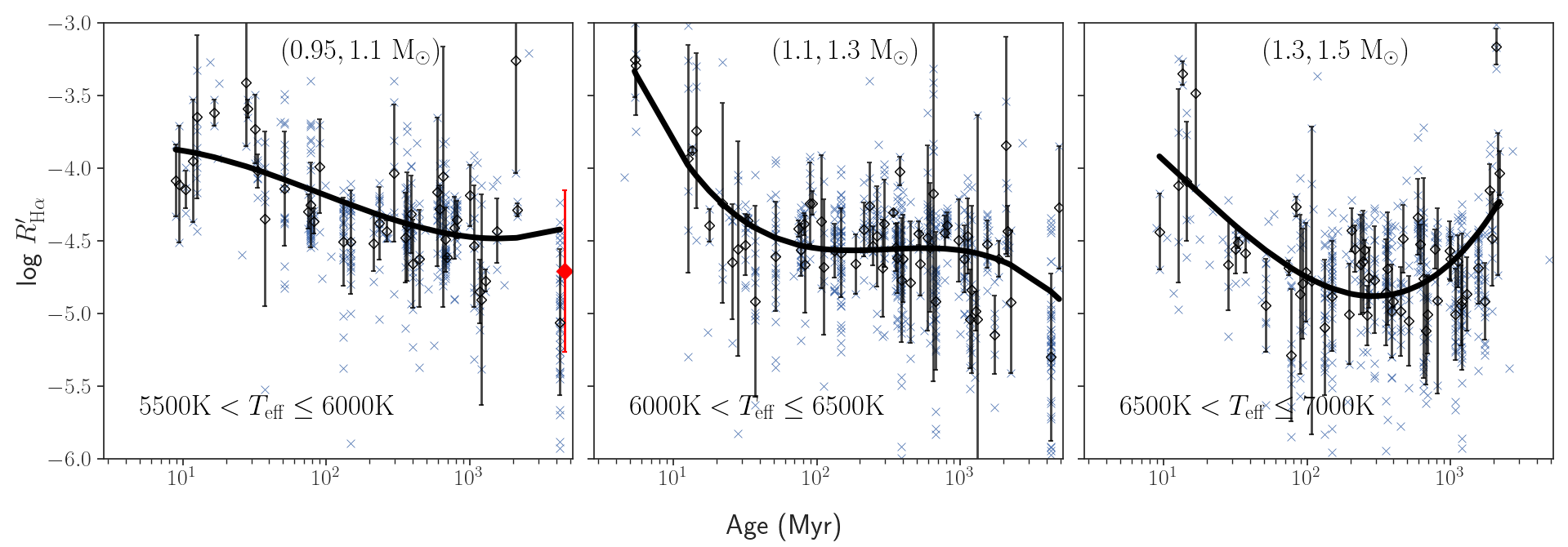}
    \caption{Chromospheric activity index, $\log R'_\mathrm{H\alpha}$, as a function of age for different effective temperature ranges. Shown are data for known star clusters with their members compiled from {\sl Gaia} DR2 by \citet{cantat2020clusters}. The data points for each star cluster have been binned, and their respective means and standard deviations are shown. The solid line represents the best-fitting fourth-order polynomial to the mean values. The red diamond corresponds to the locus of the Sun.}
    \label{fig:10}
\end{figure*}

\subsection{Core--Envelope Coupling}

Several explanations for the observed spin-down stallation effect have been discussed. One explanation suggests that bright faculae may partially cancel the effects of dark starspots \citep{reinhold2019transition,Reinhold2020}. This does not seem correct, however, since we see decreased chromospheric activity indices that are independent of the prevailing photospheric activity. Another possible explanation \citep{curtis2019temporary,spada2020competing,curtis2020stalled} is related to the process of core--envelope coupling. Core--envelope coupling assumes a two-zone approximation for stellar interiors, where the convective envelope rotates at a different rate than the radiative core. Angular momentum is assumed to be exchanged between the envelope and the core on some characteristic time-scale. In this scenario, when the age of the star is comparable to the core--envelope coupling time-scale, rotational braking of the envelope (which includes the visible surface) is temporarily stalled because the spin-down torque from the stellar wind is counteracted by a spin-up torque from the core. 

The different temperature bins in Figure \ref{fig:9} also demonstrate the efficiency of the dynamo with respect to varying convection-zone depths. Dynamo action is influenced by two key factors, i.e., rotation and convection-zone depths. Given that the range of rotation periods is similar across different temperature bins, we find it highly improbable that the difference in the convection zone alone is responsible for such significant trends. Rather, we suspect that core--envelope coupling might be the reason for the observed spin-down stallation. 

The spindown occurs due to loss of angular momentum through magnetized stellar winds. Since the chromospheric activity has decreased steeply owing to core--envelope coupling, the angular momentum loss will also be less as the stellar winds will be less magnetized. In addition, the spin-up torque from the radiative core will also slow the spin-down. Hence a combined effect leads to the spin-down stallion around the location of the Praesepe cluster. The steep decrease in activity due to coupling is possible if we consider that the regeneration of the magnetic fields happens through the dynamo interface in most stars. Conversion from poloidal to toroidal fields occurs owing to strong differential rotation at the tachocline. If core--envelope coupling occurs, it will reduce the shear and the differential rotation present at the transition of the convective envelope and the radiative core. Hence, the conversion of poloidal to toroidal fields will be reduced. This will lead to a steep decrease in magnetic activity.

We suspect that core--envelope coupling occurs effectively for K--M-type stars and that is why we observe a steep reduction in activity along the rapidly rotating sequence (see Figure \ref{fig:9}). We also observe that core--envelope coupling starts before the age of the Praesepe cluster, i.e., before an age of 670 Myr. Rather, we suspect that the radiative core and convective envelope are fully coupled till they reach the location of the period gap in the period--colour diagram, which typically occurs after 700-1000 Myr (see a schematic representation of core--envelope coupling in Figure \ref{fig:14}).  The activity--age trend in Figure \ref{fig:10}, also shows that activity decreases significantly around the age of 1 Gyr, to levels even lower than that representative of the Sun's. That is likely why we observe a steep decline in stars younger than the Praesepe cluster and a saturated, low activity level for older stars. Once the core and envelope are coupled, the activity will be minimal, since the differential rotation and the shear present at the tachocline have been reduced significantly. Therefore, K--M-type stars will remain in the minimum activity stage for a long time. This suggestion is supported by the observed, constant low activity along the slowly rotating sequence for these stars (see Figure \ref{fig:9}). F--G-type stars, meanwhile, show comparatively higher saturated activity along the slowly rotating sequence than K--M-type stars, as well as low activity levels, with a small gradient along the rapidly rotating sequence (see Figure \ref{fig:9}). We suspect that core--envelope coupling is slow in these stars, taking even longer than the age of the Sun. 

However, the double-zone model \citep{spada2020competing} suggests otherwise, i.e., that for solar-type stars core--envelope coupling takes less than 50--100 Myr. If that were the case, we suspect that core--envelope coupling is not very efficient in F--G-type stars and that is likely why we do not find much evidence of decreased activity in such stars in comparison with K--M-type stars along the slow rotating sequence. We also note a similar behaviour in the activity--age trends in Figure \ref{fig:10}, i.e., that the activity reaches a saturated level after 100 Myr and is comparatively high. That might be the reason why we observe, through helioseismology \citep{charbonneau1999helioseismic}, strong shear at the tachocline, with a differentially rotating convective zone and a solidly rotating radiative core in the Sun. We suggest that the core--envelope coupling efficiency depends strongly on the sizes and masses of the radiative core and the envelope, which is why we see core--envelope coupling occurring effectively for low-mass stars (K--M types).

\subsection{Why does the period gap occur?}

In Figure \ref{fig:8}, a gap is visible above the Praesepe cluster locus (shown as black rhombuses in Figure \ref{fig:8}). We can see the period gap ranging from 15 to 20 days and 1.3 to 2.2 mag in rotation period and {\sl Gaia} colours, respectively. It can be seen that most stars located around the period gap show saturated low-level magnetic activity (also see Figure \ref{fig:13}). In Figure \ref{fig:9}, the period gap is approximately situated where the activity levels begin to saturate and remain relatively constant (for a schematic representation, see Figure \ref{fig:14}). We suspect that the period gap arises because of the non-detection of sources in that region \citep{reinhold2019transition}. Since such stars exhibit minimal magnetic activity and do not have sufficient numbers of starspots to produce detectable photometric variability, stars located in this region remain undetected.

We suspect that around the locus of the period gap, the radiative core and convective envelope are fully coupled. Hence, minimum activity owing to a coupled core and envelope results in the non-detection of sources, thus producing the period gap. Since the core--envelope coupling time-scale is longer for M-type stars compared with K-type stars, the period gap has a positive slope in the period--colour diagram (for a schematic representation, see Figure \ref{fig:14}). Recently, \citet{lu2022bridging} found that the period gap is reduced for fully convective M-type stars. This supports our argument that the period gap is caused by the core--envelope coupling. Since fully convective stars do not have a radiative core to couple with, they do not exhibit spin-down stallation and, hence, no period gap is seen. Once the radiative core and convective envelope are fully coupled, we suspect that dynamo activity (which works at the tachocline owing to the presence of differential rotation and shear) will cease, and one possibility is that a different type of dynamo comes into action. If that is so, then the period gap may act as a transition point between both dynamo actions, hence resulting in stellar variability above the period gap.

We have also estimated the chromospheric activity indices for fully convective M-type stars (see Figure \ref{fig:11}). The data were taken from \citep{lu2022bridging}. The left-hand panel of Figure \ref{fig:11}, which shows fully convective stars, does not show a steep decrease in activity, as can be seen for semi-convective K--M-type stars (see the middle panel of Figure \ref{fig:11}). Hence, this supports our explanation that the period gap is caused by decreased activity which occurred because of core--envelope coupling.

\begin{figure*}
    \centering
    \includegraphics[width=18cm]{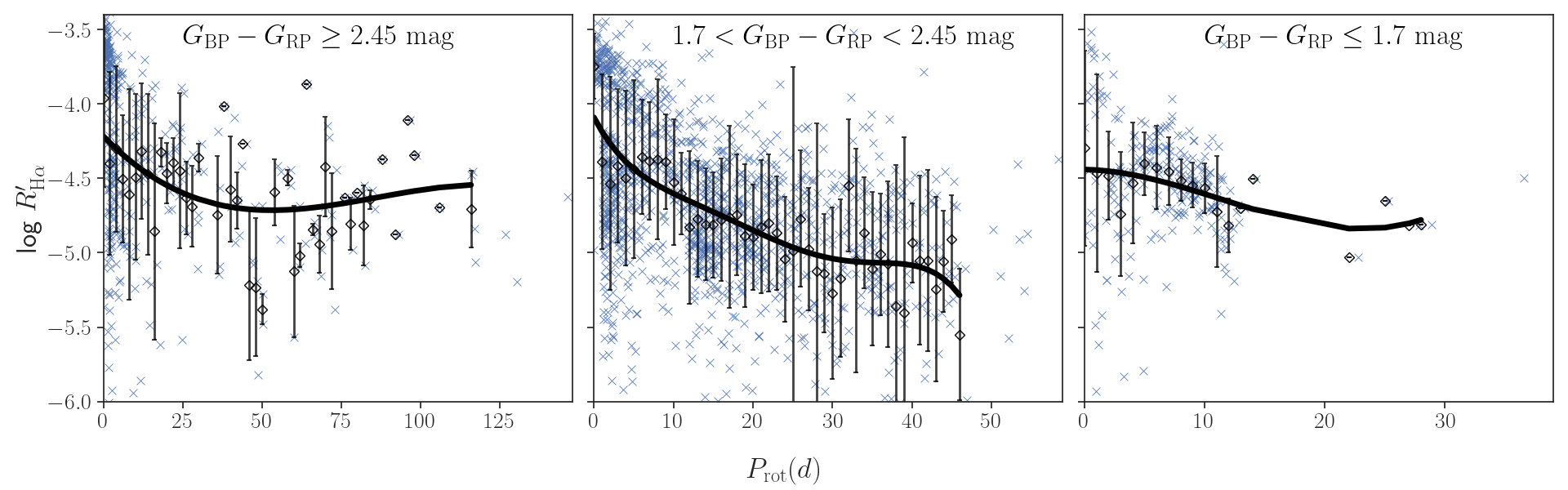}
    \caption{Chromospheric activity indices, $\log R'_\mathrm{H\alpha}$, as a function of rotation period for different temperature ranges. The data were taken from \citet{lu2022bridging}, which includes several fully convective M-type stars. The data points have been binned, and their respective means and standard deviations are also shown. The solid line represents the best-fitting fifth-order polynomial to the mean values. The left-hand panel contains fully convective M-type stars, the middle panel includes semi-convective K--M-type stars and the right-hand panel contains F--G-type stars.}
    \label{fig:11}
\end{figure*}

\section{Conclusion} \label{sec:section6}

In our previously published catalogue of BY Dra variables \citep{chahal2022statistics}, we have found traces of a period gap in the period--colour diagram. In this paper, we have investigated the origin of the period gap using chromospheric activity indices for a wide range of BY Draconis variables using LAMOST DR7 spectra. To increase our data set, we combined it with data of BY Dra variables from \citet{Mcquillan2014} and \citet{Reinhold2020}. These catalogues show evidence of a period gap. We derived Ca {\sc ii} H and K and H$\alpha$ activity indices for $\sim$15,000 and $\sim$4000 sources from, respectively, low- and medium-resolution LAMOST DR7 spectra.

We found that the chromospheric activity indices show a steep decrease until the location of the period gap, which occurs after 700--1000 Myr, followed by a constant, saturated low activity level for K--M-type stars. This provides evidence of spin-down stallation, especially in K--M-type stars. We do not observe a steep decline and a comparatively low saturated activity level for F--G-type stars. We suspect that the possible explanation for spin-down stallation is core--envelope coupling. During core--envelope coupling, after a certain period, angular momentum exchange occurs between the solid rotating radiative core and differentially rotating convective envelope. The resulting coupling will reduce the differential rotation and shear present at the tachocline, hence decreased magnetic activity. And the spin-up torque from the radiative core will lead to the spin-down stallation.

When a star reaches an age similar to that of the Praesepe cluster, we suspect that the coupling might be in the final stages. At the location of the period gap, we suspect that the radiative core and convective envelope are fully coupled. Consequently, this strong coupling results in very low magnetic activity, which may be the underlying cause of the period gap. The core--envelope coupling is effective for partially convective K--M-type stars since they have a comparable radiative core (where the central core contains most of the mass) and convective envelope. Meanwhile, for F--G-type stars, we suspect that the core--envelope coupling is weakened. In these types of stars, the convective envelope is very thin in comparison to the radiative core. That is why we observe a saturated activity level for those spectral types. 

Hence, we conclude that stars (especially K--M-type stars) show a steep decrease in activity because of core--envelope coupling which also leads to spindown stallation. We also derived activity indices for star clusters and found similar supporting evidence from the activity--age plots. We observe a steep decline in activity for K--M-type stars until an age of 1 Gyr, and a saturated activity level for F--G-type stars. \citet{lu2022bridging} showed that the period gap closes at the fully convective limit. This supports the notion that the gap is created because of core--envelope coupling since the gap is only seen for stars that have a radiative core and a convective envelope to couple.


%

\vspace{5mm}







\section*{DATA AVAILABILITY}
The LAMOST DR7 data analysed in this paper are publicly available from the LAMOST Archive. The data underlying this article are available in the article and in its online supplementary material.

\section*{Acknowledgements}
D. C. acknowledges funding support from the International Macquarie Research Excellence Scheme (iMQRES). D. K. acknowledges support from the Australian Research Council (ARC) through DECRA grant number DE190100813. This research was also supported in part by the ARC Centre of Excellence for All Sky Astrophysics in 3 Dimensions (ASTRO 3D), through project CE170100013. X. C. acknowledges funding support from the National Natural Science Foundation of China (NSFC) through grants 12173047 and 11903045. This publication is based on observations obtained with the Guoshoujing Telescope (the Large Sky Area Multi-Object Fiber Spectroscopic Telescope; LAMOST). LAMOST is a National Major Scientific Project constructed by the Chinese Academy of Sciences. Funding for the project has been provided by the Chinese National Development and Reform Commission. LAMOST is operated and managed by the National Astronomical Observatories, Chinese Academy of Sciences. We thank S. P. Rajaguru and S. Thirupathi for enlightening discussions and assistance. D. C. is grateful for the opportunity to have completed this paper as a visitor at the Indian Institute of Astrophysics. We thank T. Gordon for sharing their database. We thank the anonymous referee for providing invaluable suggestions and significantly contributing to improvements to our manuscript.




\bibliography{mnemonic,example}

\begin{thebibliography}{}
\expandafter\ifx\csname natexlab\endcsname\relax\def\natexlab#1{#1}\fi
\providecommand{\url}[1]{\href{#1}{#1}}

\bibitem[{Ag{\"u}eros {et~al.}(2018)Ag{\"u}eros, Bowsher, Bochanski, Cargile,
  Covey, Douglas, Kraus, Kundert, Law, Ahmadi, {et~al.}}]{agueros2018new}
Ag{\"u}eros, M., Bowsher, E., Bochanski, J., {et~al.} 2018, The Astrophysical
  Journal, 862, 33

\bibitem[{Angus {et~al.}(2015)Angus, Aigrain, Foreman-Mackey, \&
  McQuillan}]{angus2015calibrating}
Angus, R., Aigrain, S., Foreman-Mackey, D., \& McQuillan, A. 2015, Monthly
  Notices of the Royal Astronomical Society, 450, 1787

\bibitem[{Astudillo-Defru {et~al.}(2017)Astudillo-Defru, Delfosse, Bonfils,
  Forveille, Lovis, \& Rameau}]{Defru2017magnetic}
Astudillo-Defru, N., Delfosse, X., Bonfils, X., {et~al.} 2017, Astronomy \&
  Astrophysics, 600, A13

\bibitem[{Balona {et~al.}(2019)Balona, Handler, Chowdhury, Ozuyar, Engelbrecht,
  Mirouh, Wade, David-Uraz, \& Cantiello}]{balona2019rotational}
Balona, L., Handler, G., Chowdhury, S., {et~al.} 2019, Monthly Notices of the
  Royal Astronomical Society, 485, 3457

\bibitem[{Barnes(2007)}]{barnes2007ages}
Barnes, S.~A. 2007, The Astrophysical Journal, 669, 1167

\bibitem[{Berdyugina(2005)}]{berdyugina2005starspots}
Berdyugina, S.~V. 2005, Living Reviews in Solar Physics, 2, 1

\bibitem[{{Boro Saikia} {et~al.}(2018){Boro Saikia}, {Marvin}, {Jeffers},
  {Reiners}, {Cameron}, {Marsden}, {Petit}, {Warnecke}, \&
  {Yadav}}]{Saikia2018}
{Boro Saikia}, S., {Marvin}, C.~J., {Jeffers}, S.~V., {et~al.} 2018, A\&A, 616,
  A108

\bibitem[{Boro~Saikia {et~al.}(2018)Boro~Saikia, Marvin, Jeffers, Reiners,
  Cameron, Marsden, Petit, Warnecke, \& Yadav}]{boro2018chromospheric}
Boro~Saikia, S., Marvin, C., Jeffers, S., {et~al.} 2018, Astronomy and
  Astrophysics, 616, A108

\bibitem[{Brun {et~al.}(2015)Brun, Garc{\'\i}a, Houdek, Nandy, \&
  Pinsonneault}]{brun2015solar}
Brun, A.~S., Garc{\'\i}a, R., Houdek, G., Nandy, D., \& Pinsonneault, M. 2015,
  Space Science Reviews, 196, 303

\bibitem[{Cantat-Gaudin \& Anders(2020)}]{cantat2020clusters}
Cantat-Gaudin, T., \& Anders, F. 2020, Astronomy \& Astrophysics, 633, A99

\bibitem[{Chahal {et~al.}(2022)Chahal, de~Grijs, Kamath, \&
  Chen}]{chahal2022statistics}
Chahal, D., de~Grijs, R., Kamath, D., \& Chen, X. 2022, Monthly Notices of the
  Royal Astronomical Society, 514, 4932

\bibitem[{Charbonneau {et~al.}(1999)Charbonneau, Christensen-Dalsgaard,
  Henning, Larsen, Schou, Thompson, \& Tomczyk}]{charbonneau1999helioseismic}
Charbonneau, P., Christensen-Dalsgaard, J., Henning, R., {et~al.} 1999, The
  Astrophysical Journal, 527, 445

\bibitem[{{Cranmer} \& {Saar}(2011)}]{Cranmer2011}
{Cranmer}, S.~R., \& {Saar}, S.~H. 2011, ApJ, 741, 54

\bibitem[{Cui {et~al.}(2012)Cui, Zhao, Chu, Li, Li, Zhang, Su, Yao, Wang, Xing,
  {et~al.}}]{cui2012large}
Cui, X.-Q., Zhao, Y.-H., Chu, Y.-Q., {et~al.} 2012, Research in Astronomy and
  Astrophysics, 12, 1197

\bibitem[{Curtis {et~al.}(2019)Curtis, Ag{\"u}eros, Douglas, \&
  Meibom}]{curtis2019temporary}
Curtis, J.~L., Ag{\"u}eros, M.~A., Douglas, S.~T., \& Meibom, S. 2019, The
  Astrophysical Journal, 879, 49

\bibitem[{Curtis {et~al.}(2020)Curtis, Ag{\"u}eros, Matt, Covey, Douglas,
  Angus, Saar, Cody, Vanderburg, Law, {et~al.}}]{curtis2020stalled}
Curtis, J.~L., Ag{\"u}eros, M.~A., Matt, S.~P., {et~al.} 2020, The
  Astrophysical Journal, 904, 140

\bibitem[{da~Silva {et~al.}(2018)da~Silva, Figueira, Santos, \&
  Faria}]{da2018actin}
da~Silva, J.~G., Figueira, P., Santos, N.~C., \& Faria, J.~P. 2018, arXiv
  preprint arXiv:1811.11172

\bibitem[{da~Silva {et~al.}(2021)da~Silva, Santos, Adibekyan, Sousa, Campante,
  Figueira, Bossini, Delgado-Mena, Monteiro, de~Laverny,
  {et~al.}}]{silva2021stellar}
da~Silva, J.~G., Santos, N., Adibekyan, V., {et~al.} 2021, Astronomy \&
  Astrophysics, 646, A77

\bibitem[{David {et~al.}(2022)David, Angus, Curtis, van Saders, Colman,
  Contardo, Lu, \& Zinn}]{david2022further}
David, T.~J., Angus, R., Curtis, J.~L., {et~al.} 2022, The Astrophysical
  Journal, 933, 114

\bibitem[{{de Grijs} \& {Kamath}(2021)}]{deGrijs2021}
{de Grijs}, R., \& {Kamath}, D. 2021, Universe, 7, 440

\bibitem[{Duncan {et~al.}(1991)Duncan, Vaughan, Wilson, Preston, Frazer,
  Lanning, Misch, Mueller, Soyumer, Woodard, {et~al.}}]{duncan1991ii}
Duncan, D.~K., Vaughan, A.~H., Wilson, O.~C., {et~al.} 1991, The Astrophysical
  Journal Supplement Series, 76, 383

\bibitem[{Fang {et~al.}(2016{\natexlab{a}})Fang, Zhao, Zhao, \&
  Bharat~Kumar}]{fang2018stellar}
Fang, X.-S., Zhao, G., Zhao, J.-K., \& Bharat~Kumar, Y. 2016{\natexlab{a}},
  Monthly Notices of the Royal Astronomical Society, 476, 908

\bibitem[{Fang {et~al.}(2016{\natexlab{b}})Fang, Zhao, Zhao, Chen, \&
  Bharat~Kumar}]{fang2016stellar}
Fang, X.-S., Zhao, G., Zhao, J.-K., Chen, Y.-Q., \& Bharat~Kumar, Y.
  2016{\natexlab{b}}, Monthly Notices of the Royal Astronomical Society, 463,
  2494

\bibitem[{{Gordon} {et~al.}(2021){Gordon}, {Davenport}, {Angus},
  {Foreman-Mackey}, {Agol}, {Covey}, {Ag{\"u}eros}, \& {Kipping}}]{Gordon2021}
{Gordon}, T.~A., {Davenport}, J. R.~A., {Angus}, R., {et~al.} 2021, ApJ, 913,
  70

\bibitem[{Gray {et~al.}(2006)Gray, Corbally, Garrison, McFadden, Bubar,
  McGahee, O’Donoghue, \& Knox}]{gray2006contributions}
Gray, R., Corbally, C., Garrison, R., {et~al.} 2006, The Astronomical Journal,
  132, 161

\bibitem[{{Hall}(2008)}]{Hall2008}
{Hall}, J.~C. 2008, Living Reviews in Solar Physics, 5, 2

\bibitem[{Husser {et~al.}(2013)Husser, Wende-von Berg, Dreizler, Homeier,
  Reiners, Barman, \& Hauschildt}]{husser2013new}
Husser, T.-O., Wende-von Berg, S., Dreizler, S., {et~al.} 2013, Astronomy \&
  Astrophysics, 553, A6

\bibitem[{Li {et~al.}(2022)Li, Wang, Zeng, Liao, Du, Kong, \&
  Li}]{li2022estimation}
Li, X., Wang, Z., Zeng, S., {et~al.} 2022, Research in Astronomy and
  Astrophysics, 22, 065018

\bibitem[{Lu {et~al.}(2022)Lu, Curtis, Angus, David, \&
  Hattori}]{lu2022bridging}
Lu, Y.~L., Curtis, J.~L., Angus, R., David, T.~J., \& Hattori, S. 2022, The
  Astronomical Journal, 164, 251

\bibitem[{Mamajek \& Hillenbrand(2008)}]{mamajek2008improved}
Mamajek, E.~E., \& Hillenbrand, L.~A. 2008, The Astrophysical Journal, 687,
  1264

\bibitem[{Mascare{\~n}o {et~al.}(2016)Mascare{\~n}o, Rebolo, \&
  Hern{\'a}ndez}]{mascareno2016magnetic}
Mascare{\~n}o, A.~S., Rebolo, R., \& Hern{\'a}ndez, J.~G. 2016, Astronomy \&
  Astrophysics, 595, A12

\bibitem[{{McQuillan} {et~al.}(2014){McQuillan}, {Mazeh}, \&
  {Aigrain}}]{Mcquillan2014}
{McQuillan}, A., {Mazeh}, T., \& {Aigrain}, S. 2014, ApJS, 211, 24

\bibitem[{Montesinos {et~al.}(2001)Montesinos, Thomas, Ventura, \&
  Mazzitelli}]{montesinos2001new}
Montesinos, B., Thomas, J.~H., Ventura, P., \& Mazzitelli, I. 2001, Monthly
  Notices of the Royal Astronomical Society, 326, 877

\bibitem[{Noyes {et~al.}(1984)Noyes, Hartmann, Baliunas, Duncan, \&
  Vaughan}]{noyes1984rotation}
Noyes, R., Hartmann, L., Baliunas, S., Duncan, D., \& Vaughan, A. 1984, The
  Astrophysical Journal, 279, 763

\bibitem[{Pace(2013)}]{pace2013chromospheric}
Pace, G. 2013, Astronomy \& Astrophysics, 551, L8

\bibitem[{Parker(1993)}]{parker1993solar}
Parker, E. 1993, The Astrophysical Journal, 408, 707

\bibitem[{Reiners {et~al.}(2014)Reiners, Sch{\"u}ssler, \&
  Passegger}]{reiners2014generalized}
Reiners, A., Sch{\"u}ssler, M., \& Passegger, V. 2014, The Astrophysical
  Journal, 794, 144

\bibitem[{Reinhold {et~al.}(2019)Reinhold, Bell, Kuszlewicz, Hekker, \&
  Shapiro}]{reinhold2019transition}
Reinhold, T., Bell, K.~J., Kuszlewicz, J., Hekker, S., \& Shapiro, A.~I. 2019,
  Astronomy \& Astrophysics, 621, A21

\bibitem[{{Reinhold} \& {Hekker}(2020)}]{Reinhold2020}
{Reinhold}, T., \& {Hekker}, S. 2020, A\&A, 635, A43

\bibitem[{See {et~al.}(2021)See, Roquette, Amard, \& Matt}]{see2021photometric}
See, V., Roquette, J., Amard, L., \& Matt, S.~P. 2021, The Astrophysical
  Journal, 912, 127

\bibitem[{Spada \& Lanzafame(2020)}]{spada2020competing}
Spada, F., \& Lanzafame, A. 2020, Astronomy \& Astrophysics, 636, A76

\bibitem[{Su{\'a}rez~Mascare{\~n}o {et~al.}(2015)Su{\'a}rez~Mascare{\~n}o,
  Rebolo, Gonz{\'a}lez~Hern{\'a}ndez, \& Esposito}]{mascareno2015rotation}
Su{\'a}rez~Mascare{\~n}o, A., Rebolo, R., Gonz{\'a}lez~Hern{\'a}ndez, J., \&
  Esposito, M. 2015, Monthly Notices of the Royal Astronomical Society, 452,
  2745

\bibitem[{Van~Saders {et~al.}(2016)Van~Saders, Ceillier, Metcalfe, Aguirre,
  Pinsonneault, Garc{\'\i}a, Mathur, \& Davies}]{van2016weakened}
Van~Saders, J.~L., Ceillier, T., Metcalfe, T.~S., {et~al.} 2016, Nature, 529,
  181

\bibitem[{West {et~al.}(2004)West, Hawley, Walkowicz, Covey, Silvestri,
  Raymond, Harris, Munn, McGehee, Ivezi{\'c}, {et~al.}}]{west2004spectroscopic}
West, A.~A., Hawley, S.~L., Walkowicz, L.~M., {et~al.} 2004, The Astronomical
  Journal, 128, 426

\bibitem[{{Wilson}(1968)}]{Wilson1968}
{Wilson}, O.~C. 1968, ApJ, 153, 221

\bibitem[{Wright {et~al.}(2004)Wright, Marcy, Butler, \&
  Vogt}]{wright2004chromospheric}
Wright, J.~T., Marcy, G.~W., Butler, R.~P., \& Vogt, S.~S. 2004, The
  Astrophysical Journal Supplement Series, 152, 261

\bibitem[{Xiang {et~al.}(2015)Xiang, Liu, Yuan, Huang, Huo, Zhang, Chen, Zhang,
  Sun, Wang, {et~al.}}]{xiang2015lamost}
Xiang, M., Liu, X., Yuan, H., {et~al.} 2015, Monthly Notices of the Royal
  Astronomical Society, 448, 822

\bibitem[{Zhang {et~al.}(2019)Zhang, Zhao, Oswalt, Fang, Zhao, Liang, Ye, \&
  Zhong}]{zhang2019stellar}
Zhang, J., Zhao, J., Oswalt, T.~D., {et~al.} 2019, The Astrophysical Journal,
  887, 84

\end{thebibliography}
\bibliographystyle{aasjournal}




\section*{appendix} \label{sec:section7}

\subsection{Continuum selection}

To calculate the H$\alpha$ EW, the continuum flux was assumed to be the average flux between the spectral ranges at 6547--6557 {\AA} and 6570--6580 \AA. However, owing to the low resolution of our data, we find a significant variation in continuum flux, and hence the EWs have been altered significantly. For some sources, the EW has been reduced, while for others it increased even though they have similar line fluxes. We explored the effect of increasing the continuum ranges to 20 {\AA} and 30 {\AA} on both sides of the H$\alpha$ line, but the variation in EW persisted. To reduce the variation significantly, we applied continuum fitting to the full wavelength range of 6400--6700 \AA. Next, we derived the continuum flux using the best-fitting continuum within the wavelength ranges spanning 6527--6557 {\AA} and 6570--6600 {\AA}.

\subsection{Ca {\sc ii} H and K indices}

We observe a similar trend in the chromospheric activity--rotation plot for the Ca {\sc ii} H and K indices as we observed for the H$\alpha$ indices in Figure \ref{fig:9}. Since the FWHM of the bandpass used to calculate the $S$ index is just 1.09 {\AA}, whereas our spectra are of low resolution ($R \sim 1800$), the line emission is broadened instrumentally beyond the extent of the bandpass (see Figure \ref{fig:2}). Hence, we are not able to estimate accurate fluxes contained within the Ca {\sc ii} H and K line emission. This is not the case for H$\alpha$, since we measure its EW over a 12 {\AA} bandpass. We tried to correct for this effect by calibrating the literature-derived $S$ indices with the LAMOST $S$ indices (see Figure \ref{fig:3}). However, we still suspect that the corrected data are affected by significant uncertainties because of the low spectral resolution and possibly also since LAMOST has been measuring magnetic activity at different times (in some cases, the difference is up to two decades). Yet, we still observe similar trends with temperature for the $\log R'_\mathrm{HK}$ indices as for the $\log R'_\mathrm{H\alpha}$ indices (see Figure \ref{fig:9}). We observe a decline in activity for K--M-type stars and the activity becomes saturated around the age of the Praesepe cluster. We also observe a comparatively constant activity level for F--G-type stars. Hence, we find a reduced level of activity for both activity indices, thus providing evidence of spin-down stallation. We have also conducted a comparative analysis of the behaviour of the log $R'_{\rm HK}$ indices and the equivalent indices from \citet{noyes1984rotation,mamajek2008improved,boro2018chromospheric}. Our indices do indeed follow a similar trend as exhibited in their figures. However, it is important to acknowledge that our data set also displays significant scatter, which is primarily attributed to the wider range of stellar parameters (temperature, rotation periods, etc) included in our data.

\begin{figure*}
    \begin{multicols}{2}
    \includegraphics[width=1.05\linewidth]{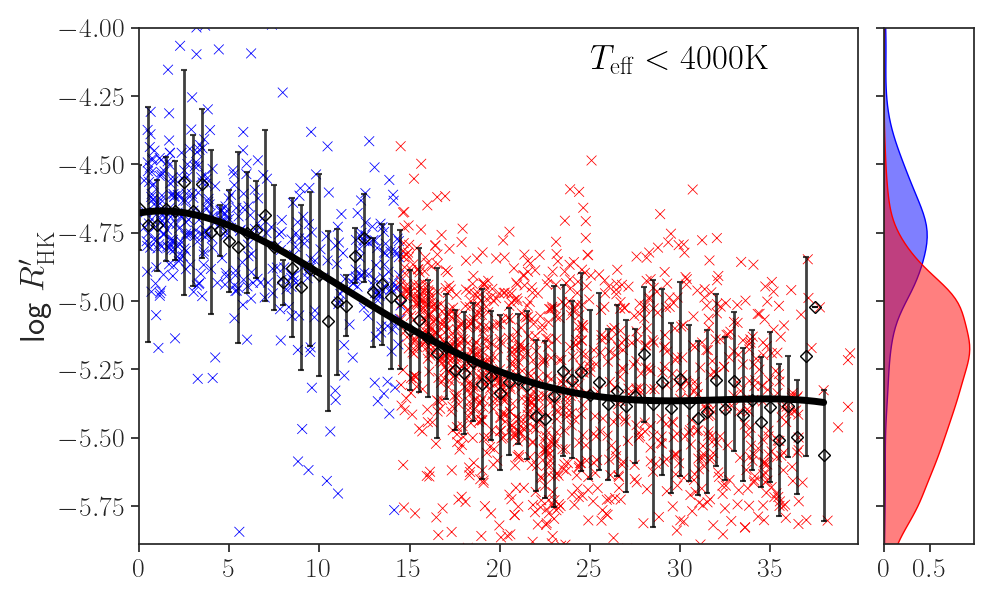}\par 
    \includegraphics[width=1\linewidth]{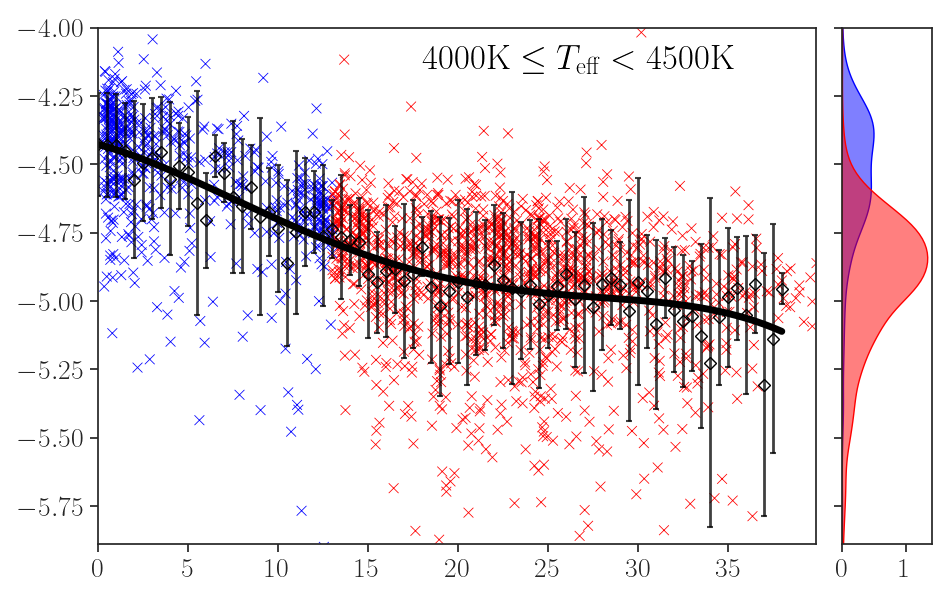}\par 
    \end{multicols}
    \begin{multicols}{2}
    \includegraphics[width=1.05\linewidth]{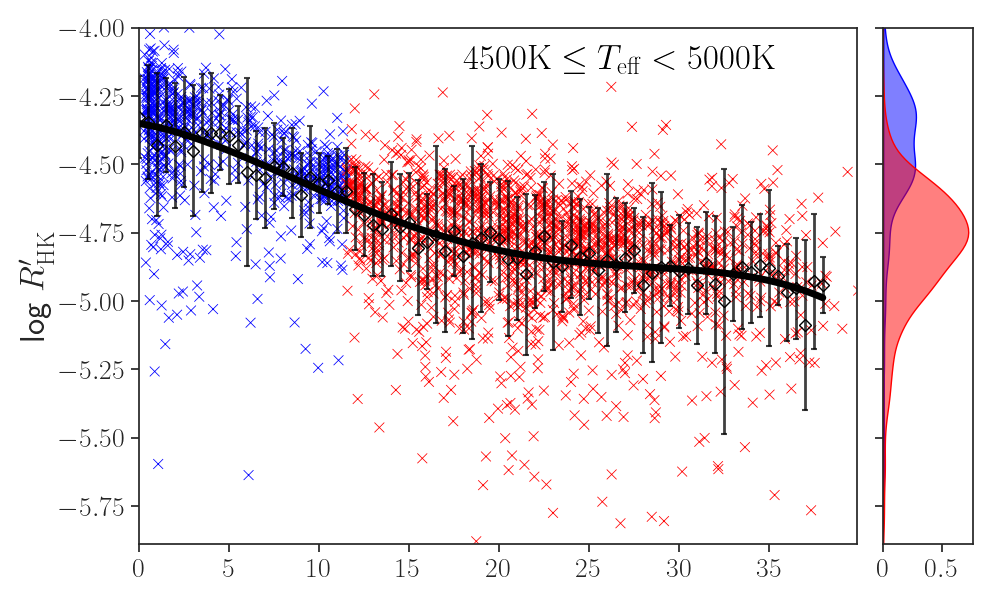}\par 
    \includegraphics[width=1\linewidth]{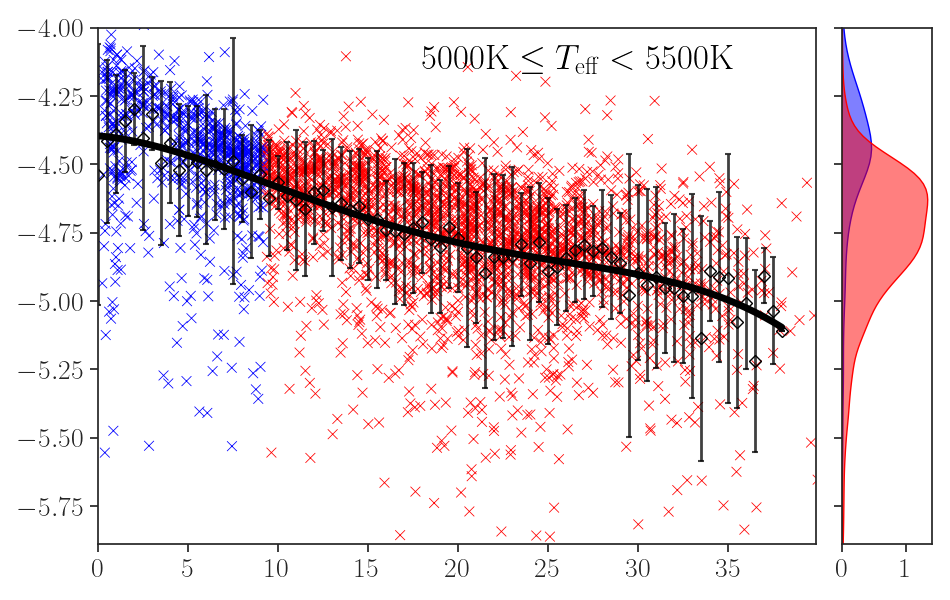}\par 
    \end{multicols}
    \begin{multicols}{2}
    \includegraphics[width=1.05\linewidth]{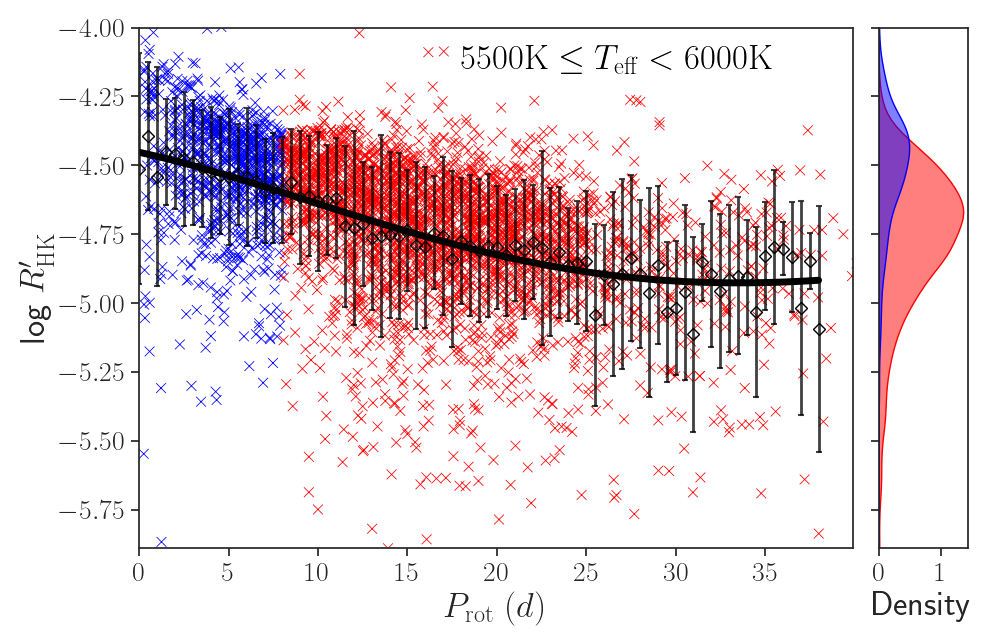}\par 
    \includegraphics[width=1\linewidth]{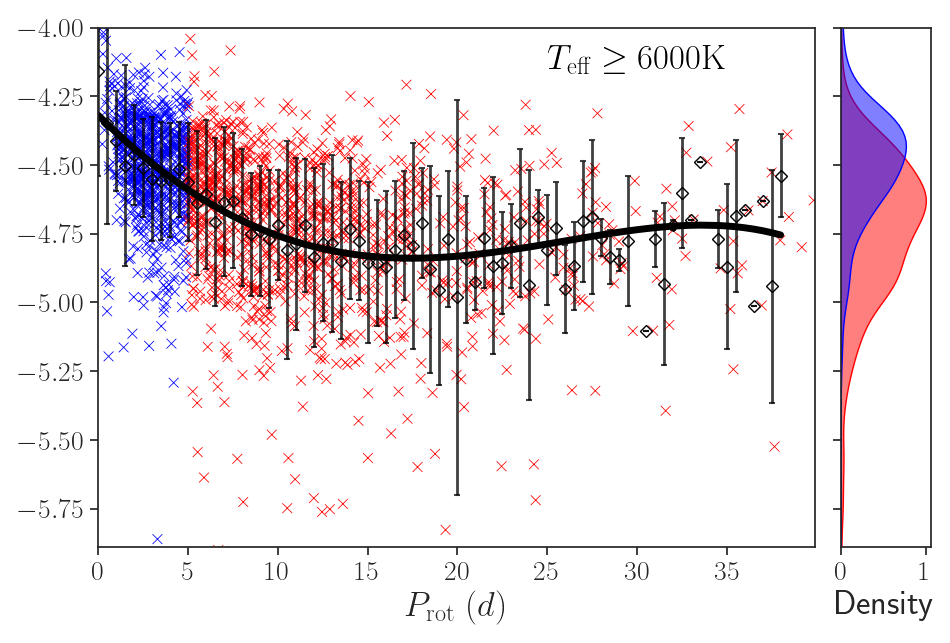}\par 
    \end{multicols}
        \caption{Chromospheric activity indices, $\log R'_\mathrm{HK}$, as a function of rotation period for different effective temperature ranges. Blue data points represent rapidly rotating stars, compared with the Praesepe cluster ($\sim$670 Myr), within their respective effective temperature ranges. Red data points are slowly rotating stars. The data points have been binned using bin widths of 0.5 days, and their respective means and standard deviations are shown. The solid line represents the best-fitting fifth-order polynomial to the mean values. The histograms on the right of each panel show the distributions of the activity indices for the blue and red data points.}
    \label{fig:12}
\end{figure*}

\begin{figure}
    \centering
    \includegraphics[width=8.8cm]{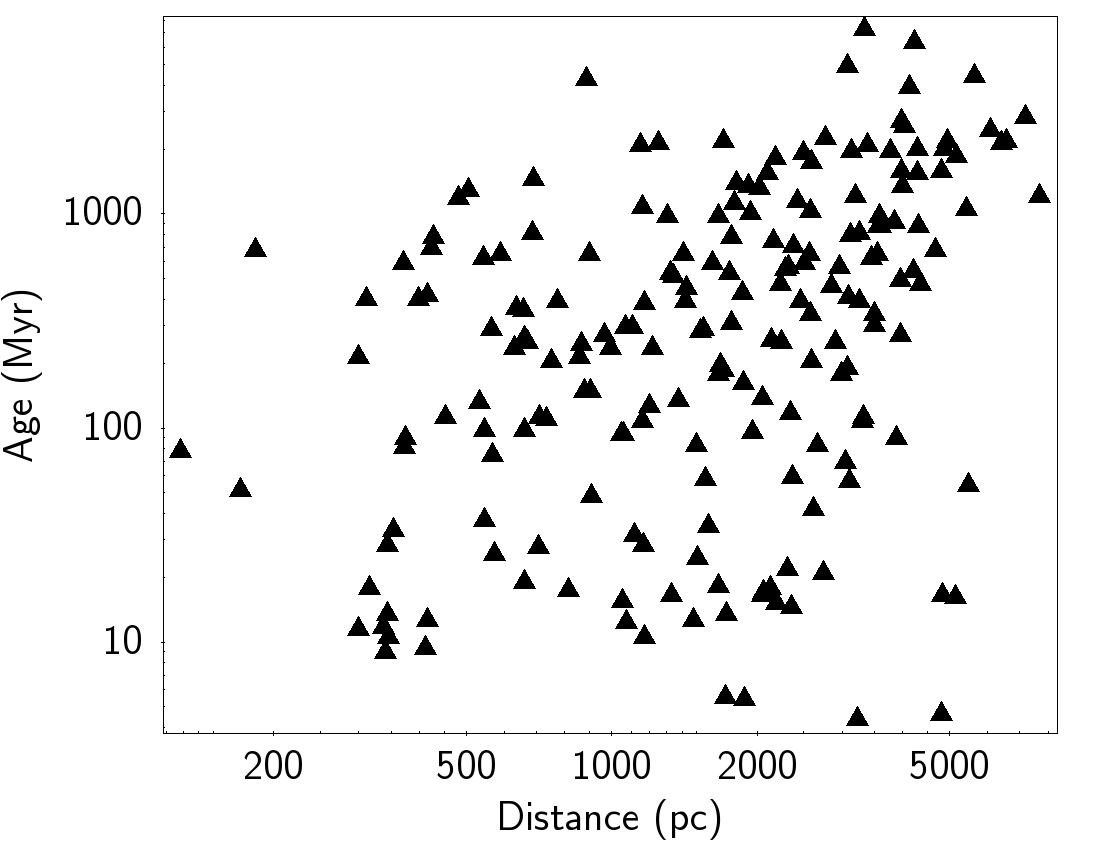}
    \caption{Age versus distances of star clusters used in our study are plotted here, taken from \citep{cantat2020clusters}. We only selected cluster members for which LAMOST DR7 spectra were available. }
    \label{fig:7}
\end{figure}



\subsection{Period gap}

\citet{Gordon2021} recently found a prominent period gap based on K2 data. These authors studied the locus of the period gap in detail. We also derived H$\alpha$ activity indices for their catalogue using the LAMOST spectra. The derived indices in the period--colour diagram are shown in Figure \ref{fig:13}. We also plotted the locus of the Praesepe cluster. We observe very low activity around the location of the period gap, as discussed in Section \ref{sec:section5}. This provides additional evidence that, independent of the data used, we see low-level activity around the period gap. Hence, our argument that the period gap may arise because of the non-detection of sources owing to very low variability levels remains valid.

\begin{figure}
    \centering
    \includegraphics[width=9cm]{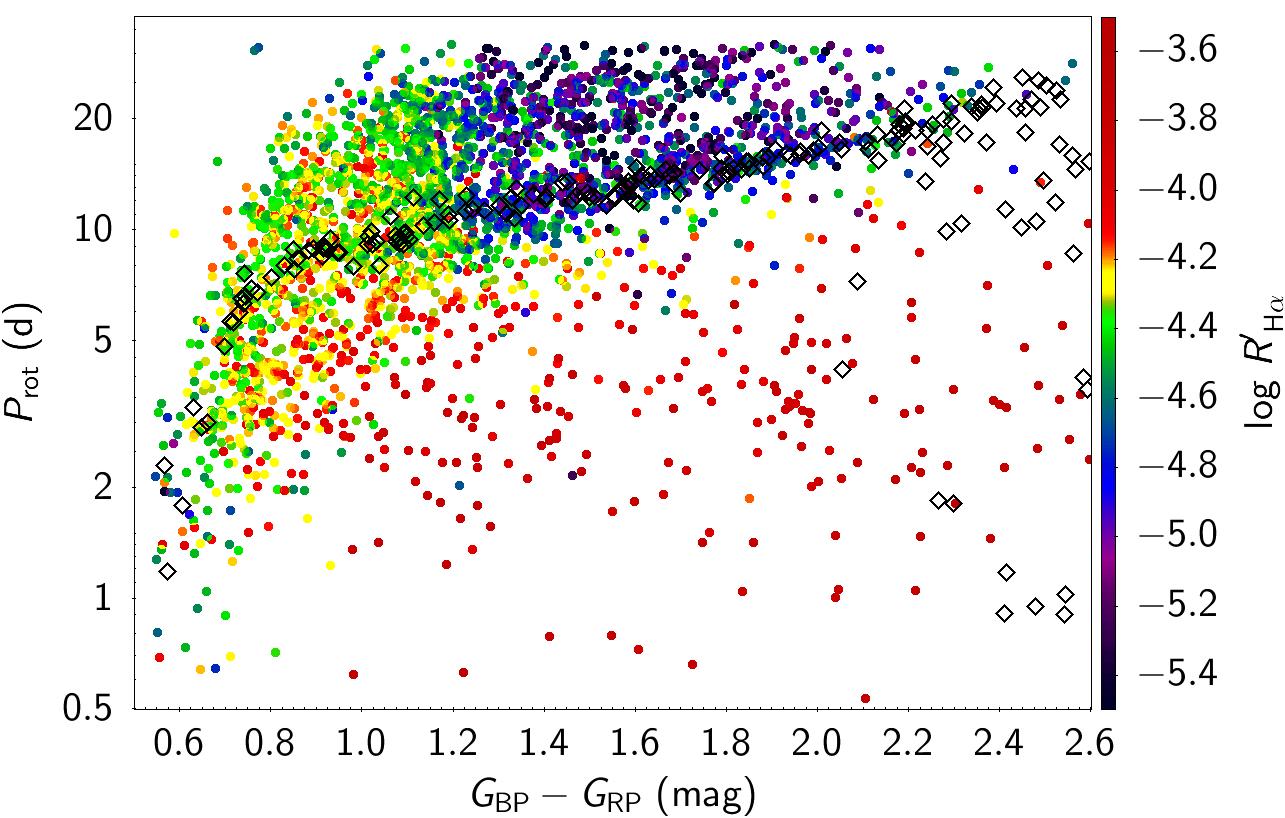}
    \caption{Rotation period as a function of {\sl Gaia} colour for the \citet{Gordon2021} data. The H$\alpha$ activity indices are colour-coded. Praesepe cluster (670 Myr) data are shown as black data points.}
    \label{fig:13}
\end{figure}

\begin{figure*}
    \centering
    \includegraphics[width=16cm]{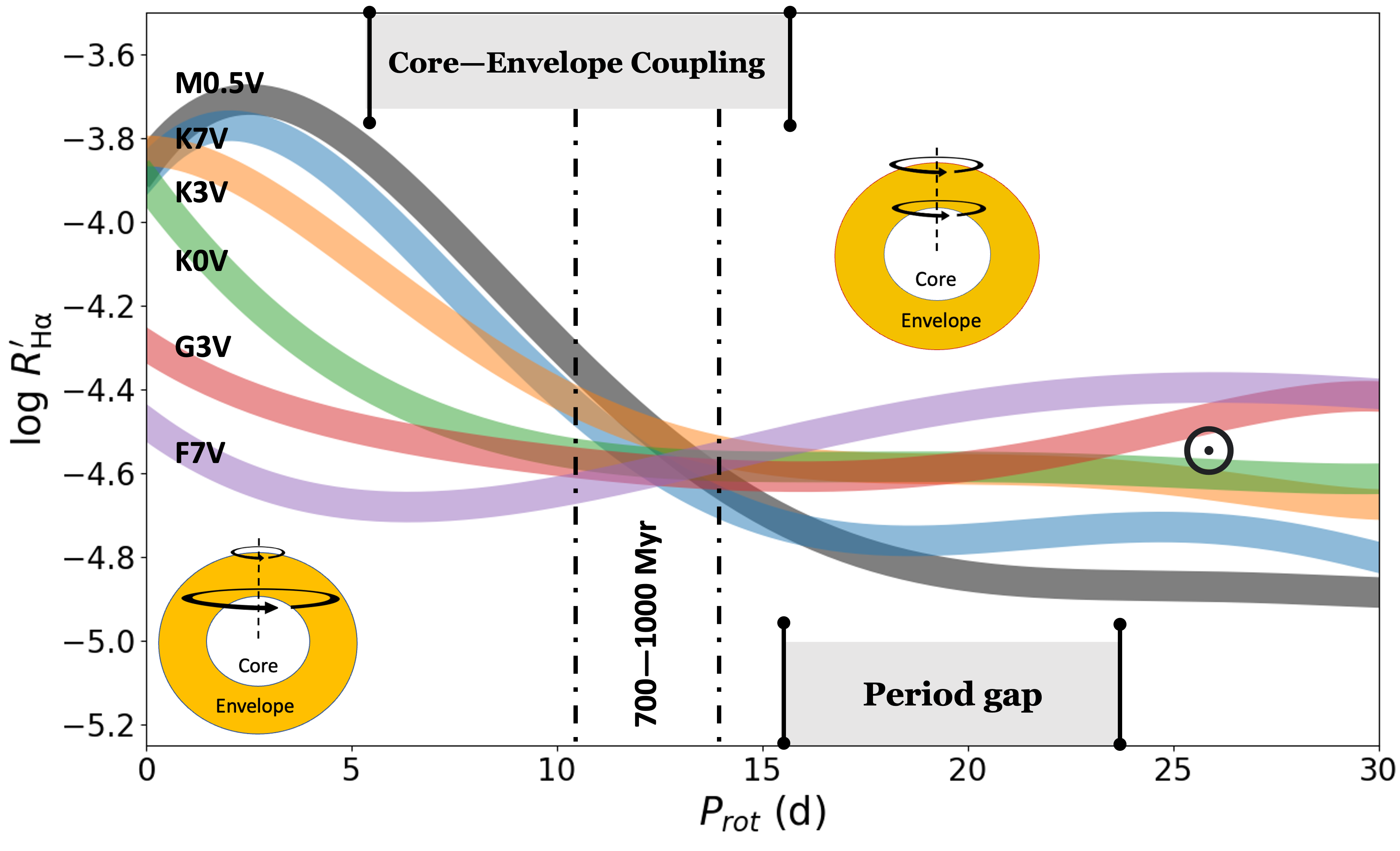}
    \caption{Schematic representation of core--envelope coupling. The polynomial fits from Figure \ref{fig:9} for different temperature bins are shown in different colours (with their respective mean spectral types labelled). The grey regions represent an average span for core--envelope coupling and the location of the period gap \citep[from][]{Gordon2021}. An illustration of stars before and after core--envelope coupling is also shown. Note that the core--envelope coupling time-scale is different for different stellar masses. Here we show an average span for K--M-type stars based on the activity--rotation trends. The larger rotating arrow represents faster rotation, and vice versa. (Not to scale.)
    }
    \label{fig:14}
\end{figure*}


\bsp	
\label{lastpage}
\end{document}